\documentclass[AMA,STIX1COL]{WileyNJD-v2}

\newcommand\BibTeX{{\rmfamily B\kern-.05em \textsc{i\kern-.025em b}\kern-.08em
T\kern-.1667em\lower.7ex\hbox{E}\kern-.125emX}}

\articletype{Article Type}%

\usepackage{times}
\usepackage{bm}
\usepackage{graphicx}
\usepackage{xcolor}
\usepackage{multirow}
\usepackage{float}
\usepackage{caption} 
\usepackage{subcaption}

\received{26 April 2016}
\revised{6 June 2016}
\accepted{6 June 2016}

\begin{document}

\title{A nonparametric relative treatment effect for direct comparisons of censored paired survival outcomes}

\author[1]{Dennis Dobler*}

\author[2]{Kathrin M\"ollenhoff}

\authormark{DENNIS DOBLER AND KATHRIN M\"OLLENHOFF}

\address[1]{\orgdiv{Department of Mathematics, Faculty of Science}, \orgname{Vrije Universiteit Amsterdam}, \orgaddress{\state{North-Holland}, \country{The Netherlands}}}

\address[2]{\orgdiv{Mathematical Institute}, \orgname{Heinrich-Heine-University D\"usseldorf}, \orgaddress{\state{North-Rhine Westphalia}, \country{Germany}}}

\corres{*Corresponding author, Vrije Universiteit Amsterdam, De Boelelaan 1111, 1081 HV, The Netherlands. \email{d.dobler@vu.nl}}

\presentaddress{Vrije Universiteit Amsterdam, De Boelelaan 1111, 1081 HV, The Netherlands}

\abstract[Summary]{A very classical problem in statistics is to test the stochastic superiority of one distribution to another. 
However, many existing approaches are developed for independent samples and, moreover, do not take censored data into account.
We develop a new estimand-driven method to compare the effectiveness of two treatments in the context of right-censored survival data with matched pairs.
With the help of competing risks techniques, the so-called relative treatment effect is estimated.
It quantifies the probability that the individual undergoing the first treatment survives the matched individual undergoing the second treatment.
Hypothesis tests and confidence intervals are based on a studentized version of the estimator, where resampling-based inference is established by means of a randomization method.
In a simulation study, we found that the developed test exhibits good power, when compared to competitors which are actually testing the simpler null hypothesis of the equality of both marginal survival functions. Finally, we apply the methodology to a well-known benchmark data set from a trial with patients suffering from with diabetic retinopathy.}


%\jnlcitation{\cname{%
%\author{D. Dobler} and
%\author{K. M\"ollenhoff}} (\cyear{2023}),
%\ctitle{A nonparametric relative treatment effect for direct comparisons of censored paired survival outcomes}, \cjournal{Stat. Med.}, \cvol{2023;00:xx--yy}.}

\maketitle

%\footnotetext{\textbf{Abbreviations:} ANA, anti-nuclear antibodies; APC, antigen-presenting cells; IRF, interferon regulatory factor}

\section{Introduction}
Testing the stochastic superiority of one distribution to another is a very classical problem in statistics.
For two independent and fully observable samples, the $t$-test, the median test, and the Mann-Whitney test are three well-known solutions to it.
Likewise, in the case of two dependent samples, i.e., if data consist of (matched) pairs, the $t$-test, the sign test, and the Wilcoxon signed rank test based on the pair-wise differences could be used.
%Extensions to the case of two dependent and \emph{censored} samples are not obvious but some have been developed in the literature:
Extensions to the case of two dependent and \emph{censored} samples are not obvious but some have been developed in the literature for a few decades. %over the past years. 
The developed test statistics are based on 
ranked absolute within-pair differences of the possibly censored survival times \citep{woolson80},
a difference of counting processes \citep{seigel82},
differences of efficient scores \citep{brien87},
ranking all censored observations separately from the uncensored ones \citep{albers88},
a combination of terms for different censoring and pair-wise ordering patterns \citep{dabrowska90},
integrals of scores with respect to differences of the sample-specific Nelson-Aalen estimators for cumulative hazards \citep{jung99},
a combination of frailty-based log-rank tests \citep{oakes10},
using further prioritized outcomes, i.e., additional data, if the primary survival endpoint does not offer decisive pair-wise comparisons \citep{pocock12}.
Extensions for covariates were also developed \citep{holt74, jeong02}. 
Reviews of methods for paired survival data including additional references to other approaches and discussions are also available\cite{woolson92,hsa14}.

While these existing approaches undoubtedly offer many good approaches for powerful statistical inference in the two-sample problem, an additional, easily interpretable quantification of the discrepancy between both samples is usually not available. 
One notable exception is the popular win ratio method of \citeauthor{pocock12}\cite{pocock12} which recently had been exploited through win odds in order to take ties into account in the inference method\cite{brunner21}.
The perhaps most straightforward approach for such a quantification is to compare the survival chances for both groups at a fixed time point.
This would provide a very limited, yet easy-to-interpret summary.
A more global impression of the difference between both samples could be obtained by integration over time, leading to the (restricted) mean survival time.
We will treat this topic in another forthcoming paper.

Instead, we will pursue an approach which is motivated by another estimand with a very clear interpretation: the \emph{relative treatment effect}.
In brief, it describes the probability that the lifetime under Treatment 1 is bigger than the lifetime under Treatment 2.
If this is (significantly) different from 0.5, a solid statistical conclusion can be drawn about the treatment efficacies.
At the same time, it is a simple probability which is easy to communicate.

In the present paper, we will develop a nonparametric methodology with an emphasis on the following quality criteria:
\begin{enumerate}
 \item begin the research with a clear formulation of an estimand of interest;
 \item make only very few and weak assumptions for the method to work;
 \item in particular, no continuity of survival functions is needed, i.e., instantaneous hazard rates need not exist;
 \item guaranteed large sample properties;
 \item a good statistical reliability even for small samples, i.e., good control of the type-I error rate and confidence level, as well as a good power and narrow confidence intervals, respectively.
\end{enumerate}
All of these points are of crucial importance, in particular in the light of  the ICH E9(R1)\cite{ICHE9R1} guidelines on estimands in trial analysis.
At the beginning of Section~A.5.1 therein, it is written that:
``An estimand for the effect of treatment relative to a control will be estimated by comparing the outcomes in a group of subjects on the treatment to those in a similar group of subjects on the control.
For a given estimand, an aligned method of analysis, or estimator, should be implemented that is able to provide an estimate on which reliable interpretation can be based. The method of analysis will also
support calculation of confidence intervals and tests for statistical significance. An important consideration for whether an interpretable estimate will be available is the extent of assumptions that
need to be made in the analysis.''

These statements clarify that even the most powerful inference method might not be the preferable one if other criteria are not met, e.g., if no intelligible estimand is available.
Also, we wish to point out that most of the methods rely on strong assumptions such as the equality of censoring times for both members of a pair\citep{woolson80, albers88} or the continuity of survival distributions\citep{woolson80, albers88, dabrowska90, oakes10}.
In this sense, the power of the test we will develop in this paper is not criterion of greatest importance although a powerful method is of course welcome.

This paper is structured as follows. First, we will introduce the relative treatment effect and explain its estimation in Section~\ref{sec:RTE}.
At the end of that section, we will relate the present approach to some others from the literature.
Second, in Section~\ref{sec:inference}, we will present a method to make statistical inference based on a data re-randomization technique\citep{dobler22}. Third, we will investigate the large sample properties of the new method and explore its small sample performance by means of a simulation study in Section~\ref{sec:sim}. Therein, also the power of the test will be assessed in a comparison to some competitor methods. Next, in Section~\ref{sec:data}, we will apply the methodology to a well-known benchmark data set from a trial with patients suffering from with diabetic retinopathy, described by \citeauthor{huster89}\cite{huster89}.
We will conclude with a discussion in Section~\ref{sec:disc}.
The Supplementary Material contains all proofs, additional technical details, and additional simulation results.

\section{Paired survival and the relative treatment effect}
\label{sec:RTE}

\subsection{Model and notation}

\noindent We denote by $(T_1, T_2)$ a bivariate random vector on a probability space $(\Omega, \mathcal A, P)$. 
Each $T_j$ $(j=1,2)$ stands for a survival time of a patient who was randomized to receive treatment $j$.
For instance, the pair results from a matching of two individuals with a similar physiology. 
As a consequence, we generally assume $T_1$ and $T_2$ to be dependent.
There exist many approaches for the estimation of the bivariate survival distribution of $(T_1,T_2)$ in the literature; see the paper by \citeauthor{pruitt93}\cite{pruitt93} for a comparison of six methods that are able to handle bivariate right-censored data;
\citeauthor{dai16}\cite{dai16} developed an estimator under the more general assumption of bivariate left-truncated and right-censored data.
We do not mean to give an exhaustive list of references in that direction and point to the references included in the two just mentioned papers.
Instead of estimating the bivariate survival function, we aim at estimating a meaningful summary of it, i.e., a treatment effect measure.

Let us now prepare the introduction of our estimand of interest.
Let $\tau >0$ denote the maximum follow-up time of a study in which the superiority of Treatment~1 to Treatment~2 shall be analyzed.
Here, superiority means that Treatment~1 prolongs the survival times compared to Treatment~2. 
Hence, it seems constructive to consider the following probability:
\begin{align}
 \label{eq:rte}
 \tilde \theta = P(T_1 > T_2) + \tfrac12 P(T_1 = T_2).
\end{align}
%In words, $\tilde p$ is the probability that the entity who received the first treatment outlives the other one who received the second treatment.
%The second term in the display is important to give equal credit to both treatments in the case of equal outcomes.
The second term is important to give equal credit to both treatments in the case of equal outcomes.
%Compared to hazard ratio analyses, the parameter $\tilde p$ seems appealing because it has a clear and direct interpretation in terms of probabilities.
Furthermore, we say that Treatment 1 is preferable if $\tilde \theta > 0.5$.
However, due to the maximum follow-up time $\tau$, $\tilde \theta$ is not always estimable; instead, we will focus on the estimand
\begin{align}
\label{eq:rte_tau}
 \theta = P(\min(T_1, \tau) > \min(T_2,\tau)) + \tfrac12 P(\min(T_1,\tau) = \min(T_2,\tau) ),
\end{align}
which we will we call the \emph{relative treatment effect} from now on.
\citeauthor{brunner00}\cite{brunner00} introduced this concept for the case of two independent samples.
Many more research papers in this context emerged afterwards. Just to mention two,
\citeauthor{munzel02}\cite{munzel02} and \citeauthor{KonietschkePauly2012}\cite{KonietschkePauly2012} considered to the case of two dependent but fully observable samples.
Finally, for technical reasons, we assume that $P(T_1 > \tau, T_2 > \tau) > 0$ and that $P(T_1 = \tau) = P(T_2 = \tau) = 0$.
The latter is always achievable by artificially increasing $\tau$ by a very small number.
Throughout the paper, we assume that $\theta \in (0,1)$, i.e., no perfect superiority of one treatment over the other.

As mentioned before, censoring is omnipresent in many medical studies.
We thus assume that survival times are independently right-censored, i.e., it is only possible to observe an event if it occurred before a so-called censoring time, say $C_1$ and $C_2$, respectively.
These are allowed to be dependent, but $(C_1, C_2)$ and $(T_1, T_2)$ are assumed to be independent.
As a consequence, the actually observable data are of the form $(X_1, \delta_1, X_2, \delta_2) $, where $X_j =\min(T_j, C_j, \tau)$ and $\delta_j = 1\{\min(T_j, \tau) \leq C_j \} $, $j=1,2$; here, $1\{\cdot \}$ denotes the indicator function.
For estimation of the second probability in the relative treatment effect, we consider the case of $X_j = \tau$ as uncensored.
On a side note, \citeauthor{efron67}\cite{efron67} also assumes the largest observation as uncensored in order to achieve a so-called ``self-consistency'' property for the Kaplan-Meier estimator; see Section~7 therein.

% \begin{rem}
	%  {For the remainder of this section, the assumption of stochastic independence of event and censoring times may be relaxed by only requiring the right-censoring to be ``independent'' in the sense that information on the censoring does not alter the compensators of counting processes for the event time; see e.g.\ Definition~III.2.1 in \cite{abgk93}.
		%  The assumptions of independent or also random censoring are often met in practice.
		%  More specifically, for paired survival outcomes, it is often natural to assume that the pair of survival times is independent of the pair of censoring times. 
		%  For instance, in the context of matching two independent patients based on similar characteristics $Z_1 = Z_2$, if one assumes conditional independence of $T_j$ and $C_j$ conditional on the sets of covariates $Z_j$, $j=1,2$, then $(T_1,T_2)$ and $(C_1,C_2)$ are also conditionally independent, although perhaps not unconditionally independent.
		%  For the latter, unconditional independence on an individual level would be sufficient.}
	% \end{rem}

In the following, we will assume without loss of generality that the censoring times are continuously distributed prior to $\tau$. In the case of discrete components in their distribution, ties can be broken by adding very small positive random numbers to them. These random numbers can be chosen small enough so that
%this does not lead to censoring times exceeding event times which had not been exceeded before. 
the order of all event times among the censoring times is not altered.
%This would reflect the continuous censoring distribution assumption.
Also, all statistical procedures considered below are not affected by these small modifications of the censoring times.

\subsection{Transformation of paired survival data into competing risks data, and estimation}
\label{ssec:trafo}

Let us now describe our novel estimation approach for the relative treatment effect.
%To this end, we will use a competing risks technique the technical details of which are described in Section~\ref{supp:cr} in the Supplementary Material.
We assume that our data set consists of $n$ independently and identically distributed data points of the kind described above, i.e., of independent paired right-censored data.
For a facilitated estimation, we transform the paired survival data into a competing risks data set. 
Note that our transformation is similar in spirit to the one used by \citeauthor{scheike14}\cite{scheike14}: therein, a transformation of a paired \emph{competing risks} data set into a univariate one facilitated the estimation of a concordance function.
Our transformation works as follows: if for a pair
\begin{itemize}
	\item[1.] the first entry is \emph{observed} to fail before the second, an event of type 1 occurred;
	\item[2.] the second entry is \emph{observed} to fail before the first, an event of type 2 occurred;
	\item[3.] both entries are \emph{observed} to fail simultaneously, an event of type 3 occurred.
\end{itemize}
All other cases are labelled right-censored.
In each case, the (censored) event time is set to be the minimum of all four event and censoring times.
We summarize the thus obtained competing risks data set as $(Z_i, \varepsilon_i) = (\min(\check T_i, \check C_i), \check \varepsilon_i \cdot  1\{\check T_i \leq  \check C_i\}),  i=1,\dots, n$.
Here, $\check T_i$ is the minimum of both (potentially unobservable) event times of pair $i$ and $\check C_i$ is the minimum of both (potentially unobservable) censoring times of pair $i$.
Note that the type of event $\check \varepsilon_i$ is unobservable in the case of a censoring.
More technical details about the transformation are given in Section~\ref{supp:cr} in the Supplementary Material.

One important consequence of the subsequent Proposition~\ref{prop:1} is that the Aalen-Johansen estimators \citep{aalen78} of $F_j$, say $\widehat F_{j,n} $, $j=1,2,3$, are available for estimating $\theta$: 
	\begin{align}
	\label{eq:rte_est}
	 \widehat \theta_n = \widehat F_{2,n}(\tau) + \tfrac12 \widehat F_{3,n}(\tau).
	\end{align}
	The specific structure of these Aalen-Johansen estimators are given in the Supplementary Material.
	Many statistical properties of the Aalen-Johansen estimator are well-known; cf.\ \cite{abgk93}, Section~IV.4.
\begin{proposition}
\label{prop:1}
\begin{enumerate}
 \item[(a)] (Representation) \ The relative treatment effect can be written as  $\theta = F_2(\tau) + \tfrac12 F_3(\tau)$,
 where $F_j(t) = P(\check T \leq t, \check \varepsilon = j)$, $t \in [0,\tau]$, denotes the $j$-th cumulative incidence function, $j=1,2,3$.
 \item[(b)] (Sufficiency) \ The above-described data reduction is sufficient for $\theta$.
 \item[(c)] (Efficiency) \ $\widehat \theta_n$ is the nonparametric maximum likelihood estimator (NPMLE) of $\theta$.
\end{enumerate}
\end{proposition}
Let us now translate two of the most important statistical results about Aalen-Johansen estimators to the relative treatment effect estimator, that is, consistency and asymptotic normality.
	To this end, we denote convergences in probability and in distribution by $\stackrel p\to$ and $\stackrel d \to$, respectively.
\begin{theorem}
	\label{thm:main}
	As $n \to \infty$, we have $\widehat \theta_n \stackrel p \to \theta$ and
	$\sqrt{n}(\widehat \theta_n - \theta) \stackrel{d}{\to} N(0,\sigma^2_\theta)$ with asymptotic variance
	$\sigma^2_\theta = \sigma_2^2 + \sigma_{23} + \tfrac14 \sigma_3^2 \in (0,\infty) $. Here, $\sigma_j^2$ and $\sigma_{jk}$, respectively, denote the asymptotic variance of $\widehat F_{j,n}$ and the asymptotic covariance of $\widehat F_{j,n}$ and $\widehat F_{k,n}$, $j,k= 1,2,3$. 
\end{theorem}
%\color{black}
%
%
%
A more detailed formula for the asymptotic variance $\sigma^2_\theta$ is offered in the Supplementary Material.
The results of Theorem~\ref{thm:main} could be combined with a consistent variance estimator to construct Wald tests and related confidence intervals.
However, such inference procedures typically exhibit a suboptimal control of the type-I error rate and the confidence level, respectively, especially for small sample sizes.
That is why we propose a resampling-based approach in Section~\ref{sec:inference} below.

\subsection{Discussion of related approaches in the literature}

Let us review the present approach in the light of existing approaches and suggestions from the literature.
\citeauthor{seigel82}\cite{seigel82} proposed to compare the counting processes for the competing risks of type 1 and 2 at time $\tau$, say $N_1(\tau) - N_2(\tau)$, i.e., McNemar's statistic. A statistical analysis of this difference would require taking the censoring rate into account.
In contrast, the relative treatment effect estimator \eqref{eq:rte_tau} relates to the difference of cumulative incidence functions through a one-to-one mapping:
$F_1(\tau) - F_2(\tau) = (1-\theta) - \theta = 1 - 2 \theta$.
Similarly, $1 - 2 \widehat \theta_n = \widehat F_{1,n}(\tau) - \widehat F_{2,n}(\tau) $. 
The latter may be written as $\int_0^\tau (N_1 - N_2)(du)/\widehat G_n(u-) $, where the denominator is the left-continuous version of the Kaplan-Meier estimator for being uncensored (in the competing risks data set).
In that sense, our approach may be called an inverse-probability-of-censoring-weighting (IPCW) version of the suggestion by \citeauthor{seigel82}\cite{seigel82}.

The previous representation also illustrates the difference from the class of test statistics suggested by \citeauthor{dabrowska90}\cite{dabrowska90}:
she proposed $T = \int K_u(s) (\tilde N_1 - \tilde N_2)(ds) + \int K_c(s) (\tilde N_3 - \tilde N_4)(ds)$.
Here $\tilde N_1$ and $\tilde N_2$ are defined like $N_1$, $N_2$, just based on the completely uncensored pairs, $\tilde N_3$ and $\tilde N_4$ correspond to the singly censored data points, and $K_u$ and $K_c$ are some scoring processes.

Let us also compare the present method to the Kaplan-Meier estimator-based approach in \citeauthor{dobler22}\cite{dobler22}.
Therein, a different variant of the relative treatment effect is analyzed, say $ \bar \theta = P(T_{11} > T_{22}) + \tfrac12 P(T_{11} = T_{22})$.
That is, outcomes from different treatments of different individuals are compared.
Other names for this parameter are \emph{D-value} or \emph{Mann-Whitney parameter}\cite{fay18,dobler18}, and also the \emph{C-index}\cite{hartman23} is related.
It is known \citep{dobler22} that $\bar \theta$ can be represented by an intergral that involves only the marginal survival functions.
Consequently, estimation could be based on two marginal and dependent Kaplan-Meier estimators.
In contrast, the method developed in the present paper is based on direct comparison within each pair. In this sense, confounding is avoided or at least reduced. The novel competing risks-based approach will thus make better use of the available information on the dependence structure within the paired observations.
It has been argued\cite{fay18,greenland20} that it is challenging to draw causal conclusions for $\bar\theta$ if not paired but only sample-specific measurements are available.
In addition, Example~\ref{ex:1} illustrates that $\bar \theta$ might exhibit some undesirable proporties which do not occur for $\theta$.
Finally, the parameter $\tilde \theta$ from \eqref{eq:rte} is related to the area under the ROC curve (AUC) which is sometimes written in a similar way.\citep{pauly16}

\begin{eexample}
\label{ex:1}
  We wish to illustrate some differences between the estimands $\theta$ and $\bar \theta$ in addition to examples from the literature\cite{fay18,greenland20}.
  In particular, we will point out two cases with different implications for subpopulations.
  Let us suppose that the following data set is fully observable; a variant of the following examples has been kindly provided by Katharina Kramer (University of Augsburg):
  \begin{center}
  \begin{tabular}{c|cc|c|c}
   pair & $T_1$ & $T_2$ & $\textnormal{sign}(T_1 - T_2)$ & subgroup \\ \hline
   1 & 2 & 1 & +1 & 1 \\
   2 & 4 & 3 & +1 & 1 \\
   3 & 6 & 5 & +1 & 2 \\
   4 & 8 & 7 & +1 & 2 \\ \hline
  \end{tabular}
  \end{center}
  Here, we also suppose that additional information (a subgroup) is available, e.g., males and females.
  For simplicity, let $\tau = \infty$.
  In the example above, the above-mentioned estimators of $\theta$ and $\bar \theta$ yield the estimates $\widehat \theta_n = 1$ and $\widehat{\bar \theta}_n = \tfrac{10}{16}$, respectively.
  These are the estimates for the whole population, i.e., both subgroups combined.
  We would like to point out that in this completely observable case the estimators simplify to the empirical fractions: $\widehat \theta_n $ is equivalent to the sign test statistic and $\widehat{\bar \theta}_n$ is equivalent to the Mann-Whitney U test statistic.
  Within the subsamples, however, the estimates are $\widehat \theta_{j,n} = 1$ and $\widehat{\bar \theta}_{j,n} = \tfrac34$, $j=1,2$.
  Extending this example reveals that $\widehat{\bar \theta}_n$ can get arbitrarily close to $\tfrac12$, i.e., basically no Treatment~1 benefit, whereas $\widehat{\bar \theta}_{j,n} =\tfrac34> \tfrac12$.
  This cannot happen with $\hat \theta_n$ which is a convex combination of the subgroup-specific $\hat \theta_{1,n}$ and $\hat \theta_{2,n}$ (in the fully observable case).
  
  In contrast, the following example is a case where $\widehat{\bar \theta}_{1,n} = \tfrac34 > \tfrac12$ and $\widehat{\bar \theta}_{2,n} = \tfrac12$ for the subsamples  but $\widehat{\bar \theta}_n = \tfrac7{16} < \tfrac12$ for the whole sample:
  \begin{center}
  \begin{tabular}{c|cc|c|c}
   pair & $T_1$ & $T_2$ & $\textnormal{sign}(T_1 - T_2)$ & subgroup \\ \hline
   1 & 2 & 1 & +1 & 1 \\
   2 & 4 & 3 & +1 & 1 \\
   3 & 5 & 6 & -1 & 2 \\
   4 & 8 & 7 & +1 & 2 \\ \hline
  \end{tabular}
  \end{center}
  That is, for each subgroup Treatment 1 seems beneficial or at least not harmful in view of $\bar \theta$, whereas is seems harmful for the whole population.
  In contrast, $\widehat \theta_n=\tfrac34$ which is in line with $\widehat \theta_{1,n} = 1$ and $\widehat \theta_{2,n} = \tfrac12$.
  This illustrates that, in some special cases, drawing conclusions for subpopulations can be difficult based on $\bar \theta$.
  As we saw, these phenomenons do not seem to occur for $\theta$.
  It should be pointed out though that such comparisons with subgroups are more challenging in the censored case because there is in general no direct connection between $\widehat \theta_n$ and $(\widehat \theta_{1,n},\widehat \theta_{2,n})$ any more.
  It might also be useful to complement the estimand $\theta$ with an average treatment effect, e.g., differences of the restricted mean survival times (RMST) $E(\min(T_1,\tau) - \min(T_2,\tau))$ to get additional insight in the effectiveness of a treatment with respect to another one.
\end{eexample}

\section{Inference on the relative treatment effect}\label{sec:inference}

One resampling option is a variant of the classical bootstrap \citep{efron1979bootstrap, Efro:cens:1981}, i.e.,  draw $n$ times independently with replacement from the competing risks data pairs $(Z_i, \varepsilon_i), i=1,\dots, n$, and recompute $\widehat \theta_n$ based on the drawn bootstrap sample.
By repeating this procedure a large number $B$ of times, the collection of the normalized bootstrapped relative treatment effect estimators, say $W_{n,b}^*  = \sqrt{n}(\widehat \theta^*_{n,b} - \widehat \theta_n), b=1, \dots, B$, can be used to estimate different aspects of the distribution of $W_n = \sqrt{n} (\widehat \theta_n - \theta)$, e.g., the $(1-\alpha)$-quantile which is necessary for a right-tailed test.
%We denote the empirical $(1-\alpha)$-quantile of $ W_{n,1}^*, \dots,  W_{n,B}^*$ by $c^*_{n,B}(1-\alpha)$, $\alpha \in (0,1)$.

Another, perhaps less well-known resampling option, is given by data re- randomization.
The general procedure is similar to the bootstrap, except that data points are not drawn with replacement but instead some other random variation is introduced which is related to an algebraic group structure.
One popular example is random permutation of the data points in a two independent samples setting which leads to permutation tests.
Randomization tests can be shown to be finitely exact if the re-randomization procedure reflects the data generation process.
Recent works revisited the finite exactness of randomization tests.\cite{hemerik18, hemerik21}
Similarly, finite exactness of confidence intervals can be shown under randomization-invariance (up to the discreteness of the randomization distribution).
In addition, \citeauthor{dobler22}\cite{dobler22} argued the asymptotic exactness of randomization tests (with finite exactness in special cases) even if the data generation process does not exactly match the re-randomization method.

We will apply one such randomization approach is the present paper.
To motivate it, consider for a moment the strong null hypothesis that both treatments are completely exchangeable in every respect.
In that case, we would have for the relative treatment effect $\theta=0.5$.
However, under the weak null hypothesis $H_0: \theta=0.5$, it is well possible that the treatments are not exchangeable; for example, if the survival functions related to $T_1$ and $T_2$ are allowed to be different.
A treatment re-randomization, i.e., a random re-labeling of the assigned treatment within each data pair, 
would lead to an artificial situation in which both treatments are exchangeable,
which in turn implies $H_0$.
That is, the relative treatment effect estimator based on the re-randomized data set targets the value $0.5$. 
Thus, randomization tests and confidence intervals that are based on critical values obtained through the described randomization procedure will be finitely exact under the sharp null hypothesis of treatment exchangeability.
Additionally, it will be asymptotically exact under the weak null hypothesis $H_0$ if the normalized relative treatment effect estimators are suitably studentized so that the limit distribution is pivotal.
Note that this re-randomization is different from the approach used in classical permutation tests for two independent samples, as the present data consist of paired data, and the randomization is done within each pair.

Let us make the randomization approach more explicit.
In terms of a fixed competing risks data set $(Z_i, \varepsilon_i), i=1,\dots,n$, a re-labeling of both treatments would mean that the times $Z_i$ remain unchanged but every occurrence of a type-1 and type-2 event will be randomly re-labeled a type-1 or type-2 event, each with probability 50\%.
Denote a thus obtained randomized data set by $(\tilde Z_i,\tilde \varepsilon_i), i=1,\dots,n$, and the resulting randomized relative treatment effect by $\tilde \theta_n$, where the randomization procedure is considered to be random.
For realizations of the re-labeling, while the original data are kept fixed, we will again use the index $b=1, \dots, B$ in the subscript to indicate the randomization iteration.
It remains to justify the conditional convergence in distribution of the randomized relative treatment effect estimator.
To this end, we introduce the notation $\mathcal L$ for the distribution of a random variable and let $d$ denote a distance which metrizes the space of distributions on $\mathbb{R}$, e.g., the Prokhorov distance\citep[pp.\ 72-73]{billingsley99}.
Finally, let $\stackrel{P}{\to}$ denote convergence in probability.
Let $\widehat \sigma_{\theta,n}^2$ and $\tilde \sigma_{\theta,n}^2$ be suitable consistent estimators of the asymptotic variances of $\widehat \theta_n $ and $ \tilde \theta_n $, respectively; see Section~\ref{supp:var} in the Supplementary Material for this paper and Subsection~\ref{ssec:simu_general} for details.
Furthermore, we wish to point out that $P\text{-}\lim_{n \to \infty} \widehat \sigma_{\theta,n}^2 = \sigma^2_\theta \neq P\text{-}\lim_{n \to \infty}  \tilde \sigma_{\theta,n}^2 =\tilde \sigma^2_\theta$ in general if the two treatments are not exchangeable.
This underlines the necessity to studentize the (randomized) estimator.
\begin{theorem}
	\label{thm:rand}
	As $n\to\infty$, 
	$$d\Big(\mathcal{L}\Big(\frac{\sqrt{n} (\tilde \theta_n - 0.5)}{\tilde \sigma_{\theta,n}} \ \Big| \ Z_1, \varepsilon_1, Z_2, \varepsilon_2, \dots \Big) ,  \mathcal{L}\Big(\frac{\sqrt{n} (\widehat \theta_n - \theta)}{\widehat \sigma_{\theta,n}}\Big) \Big) \stackrel{P}{\to} 0. $$
\end{theorem}

Based on Theorem~\ref{thm:rand},  consistent one-sided hypothesis tests for $H_0: \theta \leq 0.5$ against $H_a: \theta > 0.5$ of asymptotic level $\alpha \in (0,1)$ are given by rejecting $H_0$ if and only if $\widehat \theta_n$ exceeds $0.5 + \widehat{\sigma}_{\theta,n} \cdot c^*_{n}(1-\alpha)$. 
Here, $c^*_{n}(1-\alpha)$ denotes the conditional $(1-\alpha)$-quantile of $(\tilde \theta_n - 0.5)/\tilde\sigma_{\theta,n}$ given the data.
Such tests control the significance level exactly under the sharp null hypothesis of the exchangeability of both treatments (and censoring distributions).
Yet, they might still be slightly conservative for very small sample sizes if the non-randomized version\footnote{By a randomized version of a test, we mean tests which reject the null hypothesis with a probability $\pi \in (0,1) $ if the test statistic is equal to the critical value. This is not to be confused with the data re-randomization procedure which is used to compute the critical or $p$-value.} of the randomization tests are used.
Equivalently, $[\widehat \theta_n + \widehat{\sigma}_{\theta,n} \cdot c^*_{n}(1-\alpha) ,1]$ are one-sided asymptotic level $(1-\alpha)$ confidence intervals.
Similarly, two-sided tests and confidence intervals can be obtained if additionally the lower randomization-based quantiles are used.
As mentioned before, these confidence intervals will also be exact under treatment-exchangeability, up to the discreteness of the randomization distribution.

Another option is to apply a differentiable and one-to-one transformation $\phi: (0,1) \to \mathbb R$ to the relative treatment effect, e.g., $\phi(\theta)=\log(-\log(\theta))$.
Confidence intervals based on transformation-based statistics $\{\phi(\widehat \theta_n) - \phi(\theta)\}/\{\phi'(\widehat \theta_n) \cdot \widehat{\sigma}_{\theta,n}\}$ and the quantiles of their randomization version are ensured to be contained in $[0,1]$.
The delta-method in combination with Slutzky's theorem justifies the asymptotic correctness of these adjusted inference procedures.
% Ich glaube, hier brauchen wir doch nicht die Delta-methode zu zitieren, die ist zu basic...

%\bigskip
%\color{red}
%[Dennis an Kathrin: bioequivalence tests?? Eine Idee waere, dass man $H_0^{eq}: \theta \notin (0.5 - \varepsilon_1, 0.5 + \varepsilon_2)$ gegen $H_a^{eq}: \theta \in [0.5 - \varepsilon_1, 0.5 + \varepsilon_2]$ testen koennte...\\
%Kannst du einschaetzen, wie schwierig ein solcher Test zu entwickeln waere? Wir koennen darueber auch gerne mal am Telefon/Zoom sprechen.]
%\color{black}

\section{Simulation studies}
\label{sec:sim}

%{Achtung, laut Biometrika Styleguide ist es nicht erlaubt Text ohne subsection Zugehoerigkeit zu haben, d.h. wir muessten diesen Absatz entweder verschieben oder eine eigene subsection draus machen ODER die beiden anderen subsections aufloesen...  }
\subsection{General remarks}
\label{ssec:simu_general}

The small sample properties are analyzed with the help of simulation studies.
Next to simulations for assessing the size of the proposed right-tailed tests under the null hypothesis $H_0: \theta=0.5$, we also conducted simulations regarding the power of these tests.
All simulations were conducted under R version 4.1.0\citep{R2021}.
We also used the R package \emph{etm}\citep{allignol11} in which the Aalen-Johansen estimators and all related variance and covariance estimators {of Greenwood-type} are implemented.

\subsection{Size under the null hypothesis}

In our simulation study, we only included non-exchangeably distributed data in order to have a fair comparison between the non-randomized asymptotic, bootstrap, and randomization tests.
We considered the following combinations of simulation scenarios:
\begin{itemize}
	\item sample sizes: \ $n\in \{25,50,75,100,125,150\}$;
	\item significance levels: \ $\alpha \in \{1\%, 5\%, 10\%\} $; only the results for $\alpha=5\%$ are included in the main body of this paper, the others in Section~\ref{app:more_size} of the Supplementary Material;
	\item copulas: Gumbel-Hougaard with parameter equal to 5;\\
	Clayton with parameter equal to -0.6;
	\item marginal distributions:
	$Exp(2)$ versus a $50/50$ $Exp(3)$-$Exp(\lambda)$-mixture;
	\\
	$Gompertz(0.6,b)$ versus $Exp(3)$;
	\item censoring distributions (the same in both treatment groups): $U(0,a)$ with $a\in \{1.1, 1.6, 2.7\}$ for the first combination of marginal distributions, 
	and \\$a\in \{0.7, 1, 1.75 \}$ for the second combination;
	\item time end point $\tau = 1$ for the first combination of marginal distributions, $\tau =0.6$ for the second combination;
	\item 5,000 iterations of each tests based on 2,000 bootstrap and randomization iterations, respectively.
\end{itemize}
The above-indicated rate parameters $\lambda$ and $b$ of the marginal distributions were found through numerous generations of large data sets and they were chosen such that $\theta\approx0.5$, i.e., the null hypothesis is considered true.
The parameters of the censoring distributions resulted in censoring rates of $38\%$ to $42\%$ (strong), $27\%$ to $34\%$ (medium), and $17\%$ to $27\%$ (light); these rates were found through simulations, where truncations at $\tau$ were also considered censorings.

To elaborate a bit more on the simulation steps, we would like to point out that the bivariate copula data are generated first. Next, the quantile functions of the marginal distributions are used to transform the copula data into bivariate data with the pre-selected dependence structure and marginal distributions.
Finally, censoring is introduced by taking the minimum of the event and the simulated censoring times. 
For the purpose of estimating the relative treatment effect, the synthetic data are next transformed into a competing risks data set as described in Section~\ref{ssec:trafo}.
It should be pointed out that the cumulative incidence functions underlying the transformed competing risks data set neither have to be specified, nor do they play an important role, except for at time $\tau$.

The results displayed in Table~\ref{tab:null_5} can be summarized as follows.
Among the tests based on the untransformed relative treatment effect (``lin.'' in the table), the asymptotical test tends to be a bit liberal, while the bootstrap test is somewhat conservative.
This is more pronounced for the smaller sample sizes up to $n=100$.
The randomization test is generally closest to the selected significance level.
For larger sample sizes ($n\in \{125,150\}$), the sizes of all tests approach the  $5\%$ level quite accurately.

$$  \text{Table~\ref{tab:null_5} about here.} $$

%\begin{landscape}
%\begin{adjustbox}{angle=90}
%\Rotatebox{90}{
	\begin{sidewaystable}[h]%[!tbp]
	%\vspace{12cm}
			\caption{Simulated sizes of the right-tailed tests (in \%) with nominal significance level $\alpha=5\%$. Abbreviations: Copula: GH = Gumbel-Hougaard; critical values: asy.\ = asymptotical normal, bs.\ = bootstrap, rand.\ = randomization; tests: lin.\ = linear, tra.\ = $\log$-$\log$-transformed.}
				\centering
			\vspace{0.2cm}
		\begin{tabular}{|ccc|cc|cc|cc|cc|cc|cc|cc|cc|cc|}
\hline\noalign{\smallskip}
			%\toprule
			&&&\multicolumn{6}{c|}{light censoring}&\multicolumn{6}{c|}{medium censoring}&\multicolumn{6}{|c|}{strong censoring}\tabularnewline
			&&&\multicolumn{2}{c}{asy.}&\multicolumn{2}{c}{bs.}&\multicolumn{2}{c|}{rand.}&\multicolumn{2}{c}{asy.}&\multicolumn{2}{c}{bs.}&\multicolumn{2}{c|}{rand.}&\multicolumn{2}{c}{asy.}&\multicolumn{2}{c}{bs.}&\multicolumn{2}{c|}{rand.}\tabularnewline
			copula & distribution & $n$ & lin. & tra. &lin. & tra. &lin. & tra. &lin. & tra. &lin. & tra. &lin. & tra. &lin. & tra. &lin. & tra. &lin. & tra.  \tabularnewline
			%\midrule
			\hline
			GH&Exp vs.&$ 25$&$7.0$&$4.4$&$3.7$&$5.8$&$5.6$&$5.5$&$7.1$&$4.8$&$4.3$&$6.5$&$5.7$&$5.6$&$ 6.4$&$4.5$&$3.6$&$5.1$&$ 6.7$&$ 6.8$\tabularnewline
			&Exp mix&$ 50$&$6.0$&$4.4$&$4.3$&$5.2$&$5.2$&$5.2$&$6.2$&$4.7$&$4.4$&$5.4$&$5.4$&$5.4$&$ 5.2$&$4.2$&$3.3$&$3.9$&$ 6.2$&$ 6.2$\tabularnewline
			&&$ 75$&$5.6$&$4.3$&$4.3$&$5.0$&$5.0$&$5.0$&$6.7$&$5.2$&$4.9$&$5.8$&$5.9$&$5.9$&$ 4.5$&$3.8$&$3.0$&$3.5$&$ 5.9$&$ 5.8$\tabularnewline
			&&$100$&$5.5$&$4.6$&$4.7$&$5.2$&$5.2$&$5.2$&$5.9$&$4.7$&$4.6$&$5.2$&$5.1$&$5.1$&$ 4.9$&$4.2$&$3.3$&$3.8$&$ 6.8$&$ 6.8$\tabularnewline
			&&$125$&$5.1$&$4.3$&$4.5$&$4.9$&$4.9$&$4.9$&$5.4$&$4.5$&$4.5$&$4.8$&$5.1$&$5.1$&$ 4.7$&$4.3$&$3.5$&$3.7$&$ 7.2$&$ 7.2$\tabularnewline
			&&$150$&$5.8$&$5.1$&$5.3$&$5.5$&$5.5$&$5.5$&$5.7$&$4.9$&$5.0$&$5.4$&$5.3$&$5.3$&$ 4.5$&$3.9$&$3.5$&$3.7$&$ 6.7$&$ 6.7$\tabularnewline \hline
			GH&Gompertz&$ 25$&$8.1$&$5.0$&$2.2$&$4.5$&$6.2$&$6.2$&$9.7$&$7.0$&$3.0$&$5.8$&$9.3$&$9.5$&$11.5$&$8.8$&$4.2$&$5.1$&$15.9$&$15.5$\tabularnewline
			&vs.\ Exp&$ 50$&$6.0$&$4.3$&$3.7$&$4.6$&$5.2$&$5.2$&$6.2$&$4.5$&$2.6$&$3.6$&$5.9$&$5.9$&$ 8.4$&$6.6$&$2.8$&$4.4$&$12.7$&$12.7$\tabularnewline
			&&$ 75$&$5.0$&$3.7$&$3.6$&$4.4$&$4.2$&$4.2$&$5.3$&$4.0$&$3.0$&$3.8$&$5.0$&$5.1$&$ 7.1$&$5.8$&$2.8$&$3.9$&$12.2$&$12.2$\tabularnewline
			&&$100$&$5.3$&$4.3$&$4.5$&$5.0$&$5.0$&$5.0$&$5.8$&$4.5$&$4.0$&$4.6$&$5.4$&$5.4$&$ 5.7$&$4.4$&$2.3$&$2.8$&$10.1$&$10.1$\tabularnewline
			&&$125$&$5.4$&$4.5$&$4.6$&$5.1$&$5.2$&$5.2$&$5.5$&$4.6$&$4.6$&$5.0$&$5.2$&$5.2$&$ 5.0$&$4.1$&$2.4$&$2.9$&$ 9.8$&$ 9.8$\tabularnewline
			&&$150$&$5.1$&$4.4$&$4.6$&$4.9$&$4.8$&$4.8$&$5.3$&$4.3$&$4.5$&$4.9$&$4.8$&$4.8$&$ 4.4$&$3.9$&$2.6$&$2.9$&$ 9.1$&$ 9.1$\tabularnewline\hline
			Clayton&Exp vs.&$ 25$&$6.2$&$4.2$&$3.2$&$5.2$&$5.0$&$5.0$&$6.4$&$3.9$&$3.1$&$5.4$&$5.2$&$5.2$&$ 6.4$&$4.0$&$3.1$&$5.6$&$ 4.7$&$ 4.7$\tabularnewline
			& Exp mix&$ 50$&$5.5$&$4.0$&$3.8$&$4.7$&$4.8$&$4.8$&$5.9$&$4.3$&$3.7$&$4.7$&$4.9$&$4.9$&$ 5.6$&$4.0$&$3.4$&$4.6$&$ 5.4$&$ 5.4$\tabularnewline
			&&$ 75$&$5.6$&$4.5$&$4.3$&$4.9$&$5.0$&$5.0$&$5.3$&$4.2$&$3.9$&$4.5$&$4.8$&$4.8$&$ 5.3$&$4.4$&$4.0$&$4.6$&$ 5.4$&$ 5.4$\tabularnewline
			&&$100$&$5.5$&$4.6$&$4.6$&$4.9$&$5.4$&$5.4$&$5.3$&$4.3$&$4.2$&$4.6$&$4.8$&$4.8$&$ 5.0$&$4.1$&$3.7$&$4.1$&$ 5.0$&$ 4.9$\tabularnewline
			&&$125$&$4.7$&$3.9$&$4.0$&$4.3$&$4.7$&$4.7$&$5.3$&$4.7$&$4.7$&$5.0$&$5.3$&$5.3$&$ 5.2$&$4.4$&$3.8$&$4.2$&$ 5.3$&$ 5.3$\tabularnewline
			&&$150$&$4.9$&$4.1$&$4.3$&$4.5$&$4.6$&$4.6$&$4.9$&$4.1$&$4.3$&$4.5$&$4.9$&$4.9$&$ 5.1$&$4.6$&$4.0$&$4.2$&$ 5.4$&$ 5.4$\tabularnewline\hline
			Clayton&Gompertz&$ 25$&$6.8$&$4.5$&$3.0$&$5.5$&$5.4$&$5.4$&$7.3$&$4.5$&$3.3$&$5.9$&$5.3$&$5.3$&$ 8.3$&$5.1$&$3.9$&$6.7$&$ 5.8$&$ 5.8$\tabularnewline
			& vs.\ Exp&$ 50$&$5.5$&$4.0$&$3.6$&$4.6$&$5.0$&$5.0$&$6.7$&$5.0$&$4.1$&$5.2$&$5.8$&$5.8$&$ 6.7$&$4.9$&$3.9$&$5.2$&$ 5.6$&$ 5.6$\tabularnewline
			&&$ 75$&$5.4$&$4.2$&$4.0$&$4.8$&$4.9$&$4.9$&$6.0$&$4.3$&$4.0$&$4.9$&$5.4$&$5.4$&$ 6.0$&$4.6$&$3.8$&$4.7$&$ 5.5$&$ 5.5$\tabularnewline
			&&$100$&$5.8$&$4.8$&$4.8$&$5.2$&$5.4$&$5.4$&$5.7$&$4.7$&$4.3$&$4.7$&$5.3$&$5.3$&$ 6.3$&$5.0$&$4.2$&$4.8$&$ 5.8$&$ 5.8$\tabularnewline
			&&$125$&$4.8$&$4.0$&$4.0$&$4.4$&$4.6$&$4.6$&$5.1$&$4.5$&$4.2$&$4.6$&$4.9$&$4.9$&$ 5.9$&$4.9$&$4.3$&$4.8$&$ 5.8$&$ 5.8$\tabularnewline
			&&$150$&$5.4$&$4.7$&$4.7$&$5.1$&$5.2$&$5.2$&$5.9$&$4.8$&$4.7$&$5.0$&$5.5$&$5.5$&$ 5.7$&$4.9$&$4.4$&$4.8$&$ 5.6$&$ 5.6$\tabularnewline
			%\bottomrule
			\hline
		\end{tabular}
		\label{tab:null_5}
	\end{sidewaystable}
	%}
%\end{landscape}
%\end{adjustbox}

\begin{figure}[ht]
\includegraphics[width=0.45\textwidth]{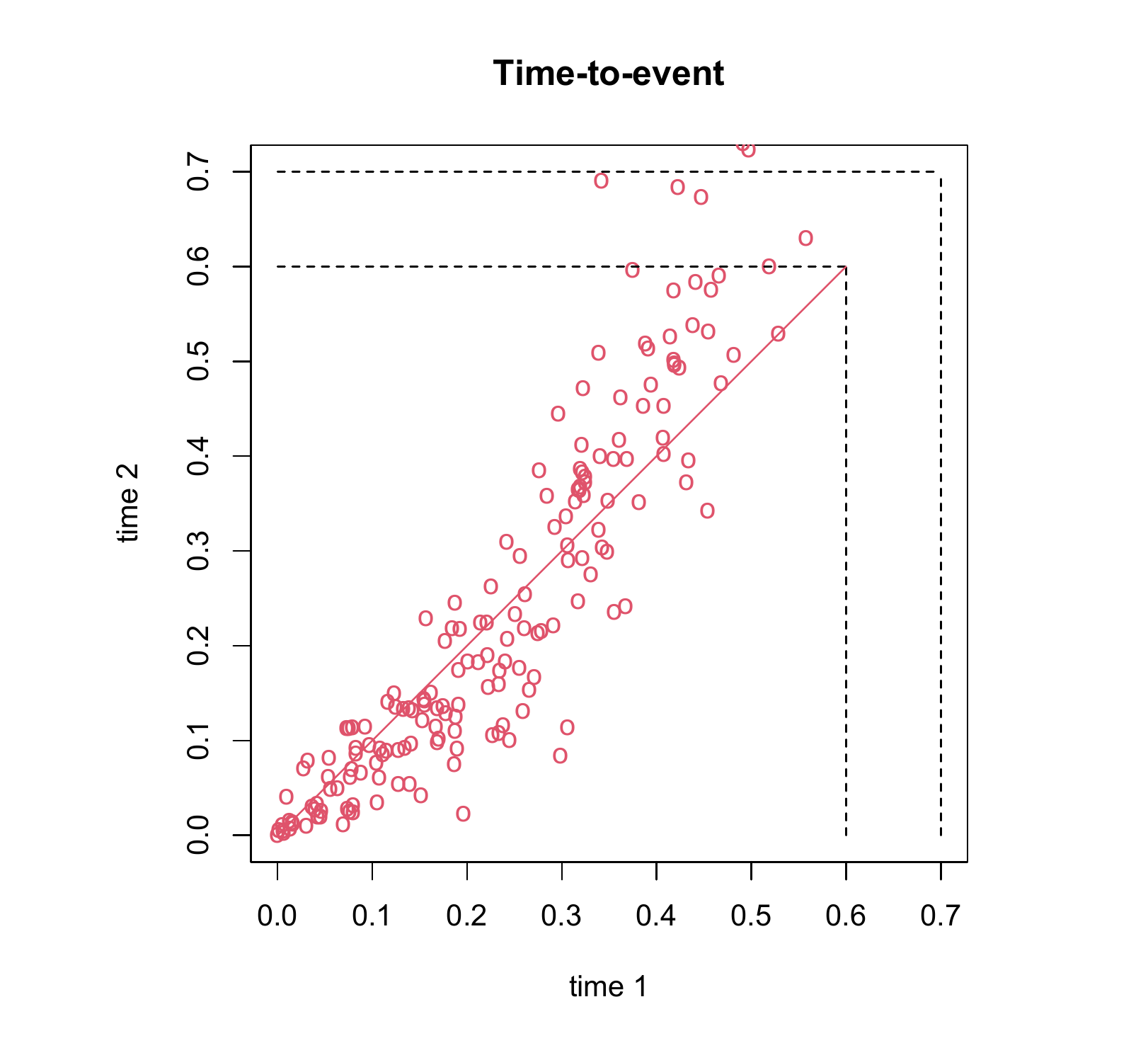}
\
\includegraphics[width=0.45\textwidth]{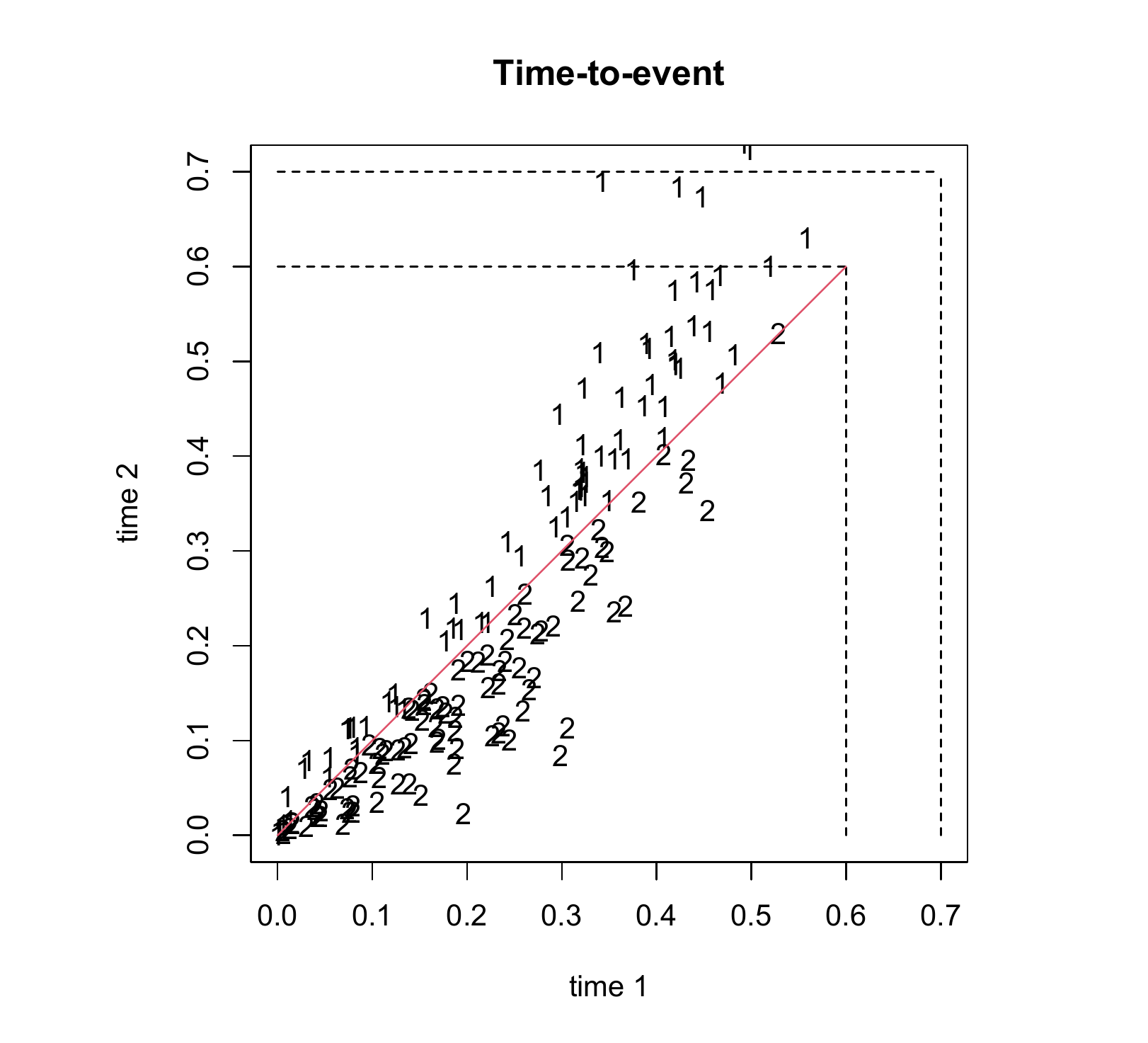}
 \\
\includegraphics[width=0.45\textwidth]{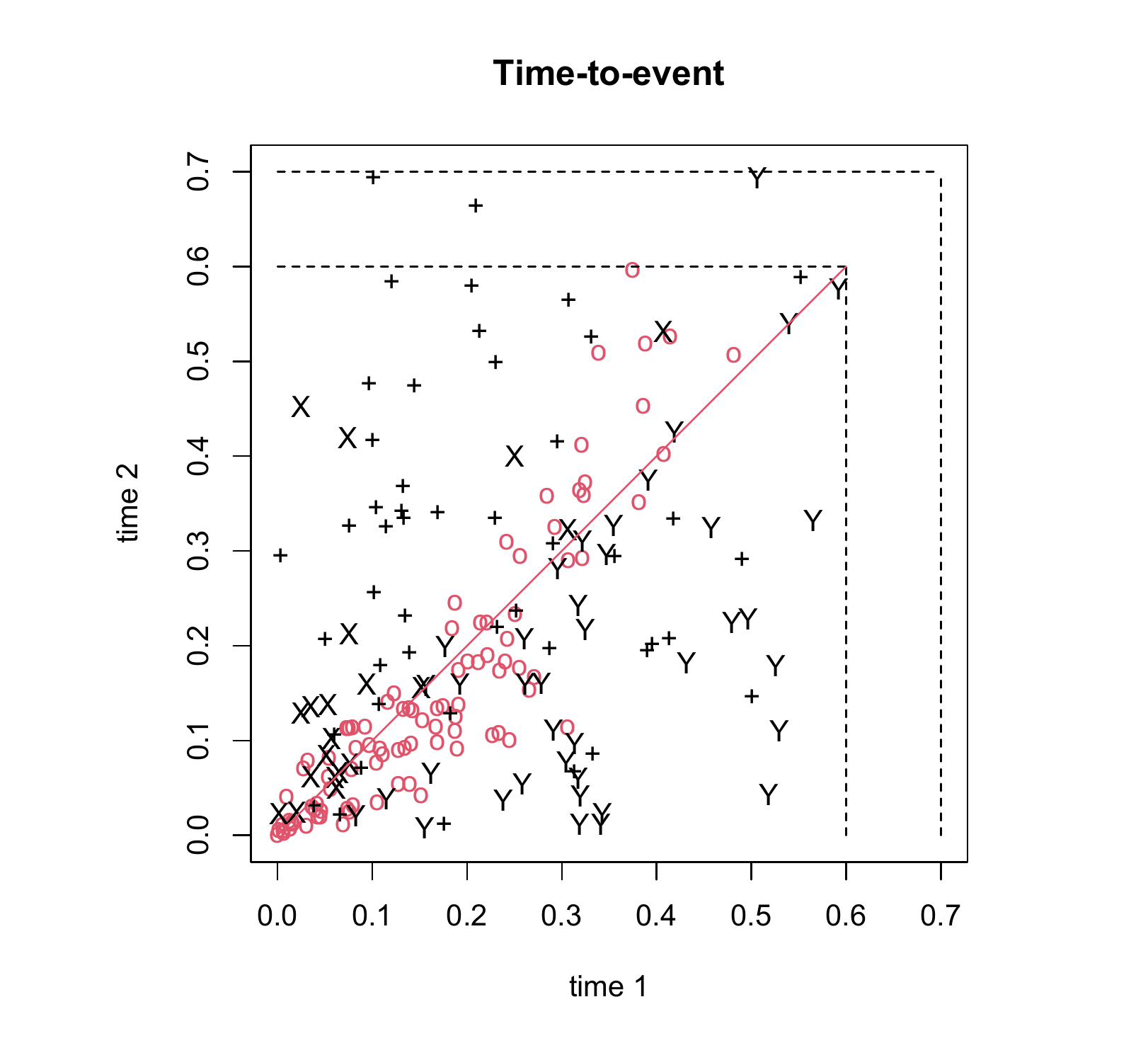}
\
\includegraphics[width=0.45\textwidth]{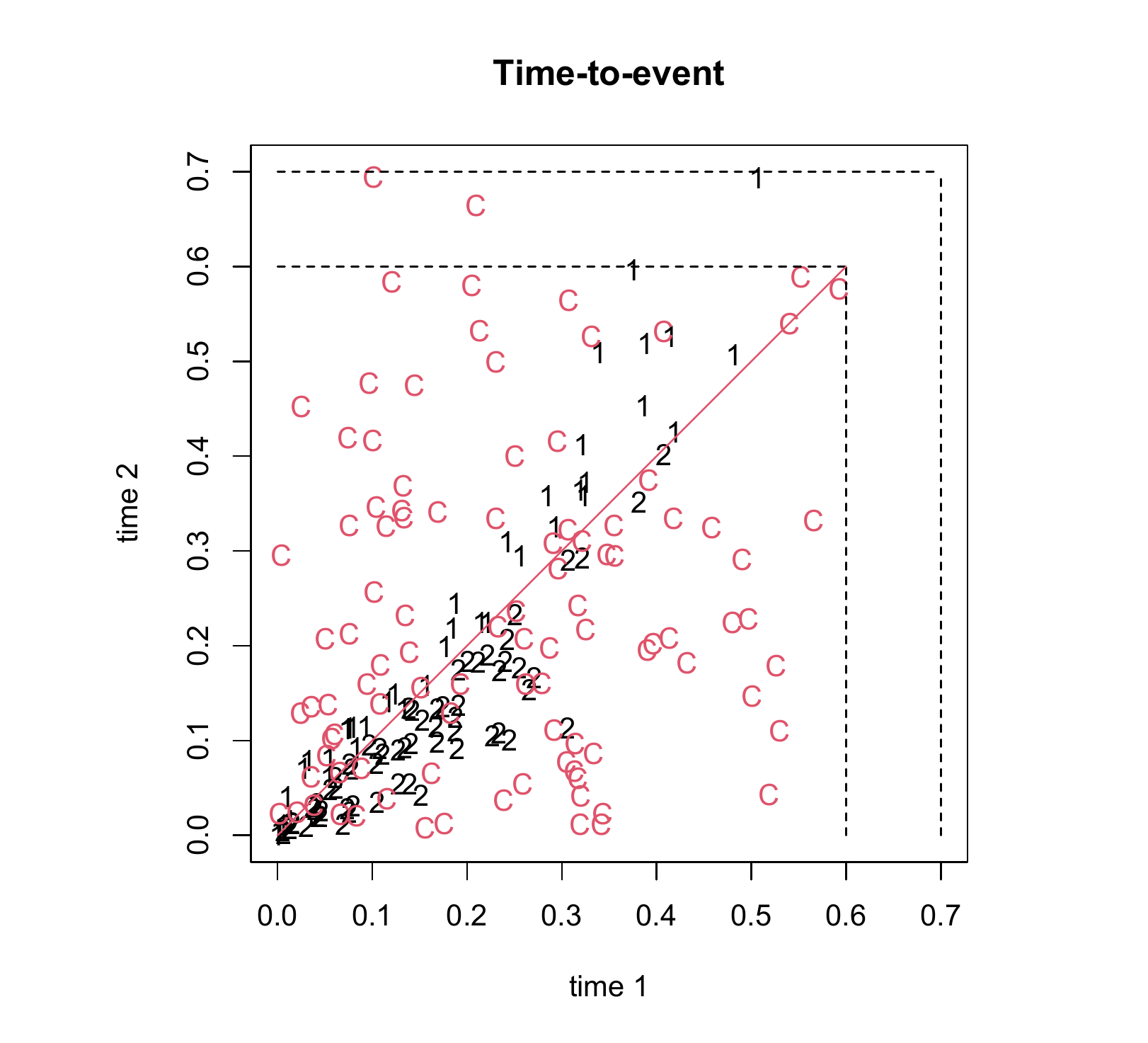}
 \caption{Scatterplots of $n=200$ simulated data points according to a Gumbel-Hougaard copula and, respectively, Gumbel and exponentially distributed marginals. 
 The red line is the diagonal $y=x$.
 The dashed lines illustrate the upper end of the censoring support (0.7) and $\tau=0.6$.
 \\
 Upper panels: complete data set (no censoring); lower panels: censored data set.
 \\
 %Note: the maximum of the horizontal times is 0.68; the maximum of the vertical times is 1.41.\\
 Left panels: raw data; censorings in both coordinates are denoted by "+"; censorings only in the horizontal coordinate are denoted "X"; censorings only in the vertical coordinate are denoted "Y"; completely uncensored data points are denoted by red circles.
 \\
 Right panels: competing risks data after transformation.
 The event time is the minimum of both coordinates. 
 The symbols "1", "2", and "C" represent whether the data point corresponds to an observed event of type 1, type 2, or to a censoring.}
 \label{fig:GH_scatter}
\end{figure}

One notable peculiarity is the combination of strong censoring, the Gumbel-Hougaard copula, and, in particular, Gompertz versus exponential marginals: here, the bootstrap test stays very conservative, and the randomization test is very anti-conservative. 
There is a slight improvement when the sample size increases.
In this scenario, the asymptotic test is also anti-conservative but it surprisingly performs better than the randomization test.
For smaller sample sizes ($n\in \{25,50\}$), similar observations about the simulation results can be made for the related medium censoring case.
Figure~\ref{fig:GH_scatter} illustrates that the above-discussed scenario is indeed a very challenging one: many data points are converted into censorings on the competing risks scale; the censoring rate amounts to about 52\%.
At the same time, there is a very strong correlation between the survival times of a pair, and two very different marginal distributions.
For smaller sample sizes, most of these characteristics are hardly visible.
That is why this is by far the most challenging simulation setting.

The $\log$-$\log$-transformation (``tra.'' in the table) seems to rectify the liberality of the asymptotic test and the conservativeness of the bootstrap test in most scenarios.
However, the transformation has nearly no effect on the randomization test.

\subsection{Power simulations}

In addition to the size simulations under $H_0: \theta=0.5$, we have also conducted a simulation study to assess the power of the one-tailed version of the developed test.
Since the results for the transformed and untransformed test statistics are very much alike, we have solely focused on the latter.
We also considered two competitor tests, also in their one-tailed versions: the paired Prentice-Wilcoxon test\cite{brien87}
%which is based on an efficient scores approach and 
which was found to be very powerful in the comparative simulation study by \cite{woolson92}; 
the stratified log-rank test\cite{oakes10} which was more closely analyzed under correlated frailty models.
In order to ensure a fair comparison, the randomization versions of all tests were used, such that all of them control the significance level for finite sample sizes under exchangeability.
In this subsection, we chose the significance level $\alpha=5\%$.

We considered the same two copulas as in the previous subsection, the sample sizes $n \in \{25,50,100\}$, 1,000 test replications, 1,000 randomization iterations, and the following three marginal distribution scenarios:
\begin{enumerate}
	\item Mixture of the $Exp(2)$-exponential and the $U(0,2)$-uniform distribution against the $Exp(2)$-exponential distribution; the censoring times were independently\\ $U(0,2.5)$-distributed and $\tau = 1.9$. This scenario departs from the sharp null hypothesis of exchangeability into alternatives with crossing hazard rates at late time points close to $\tau$  as the mixing parameter puts more and more weight on the uniform distribution.
	\item Mixture of the $Exp(2)$-exponential and the Gompertz distribution with shape parameter $0.1$ and rate parameter $2$ against the $Exp(2)$-exponential distribution; the censoring times were independently $U(0,2.5)$-distributed and $\tau = 1.8$. This scenario departs from the sharp null hypothesis of exchangeability into alternatives with crossing hazard rates at central time points as the mixing parameter puts more and more weight on the Gompertz distribution.
	\item The $Exp(2/k)$-exponential distribution  against the $Exp(2)$-exponential distribution; the censoring times were independently $U(0,2)$-distributed and $\tau = 1.3$. This scenario departs from the sharp null hypothesis of exchangeability into alternatives with parallel hazard rates as the scale parameter $k$ increases from $1$ to $2$.
\end{enumerate}

\begin{figure}[ht]
	\centering
	\includegraphics[width=1\textwidth]{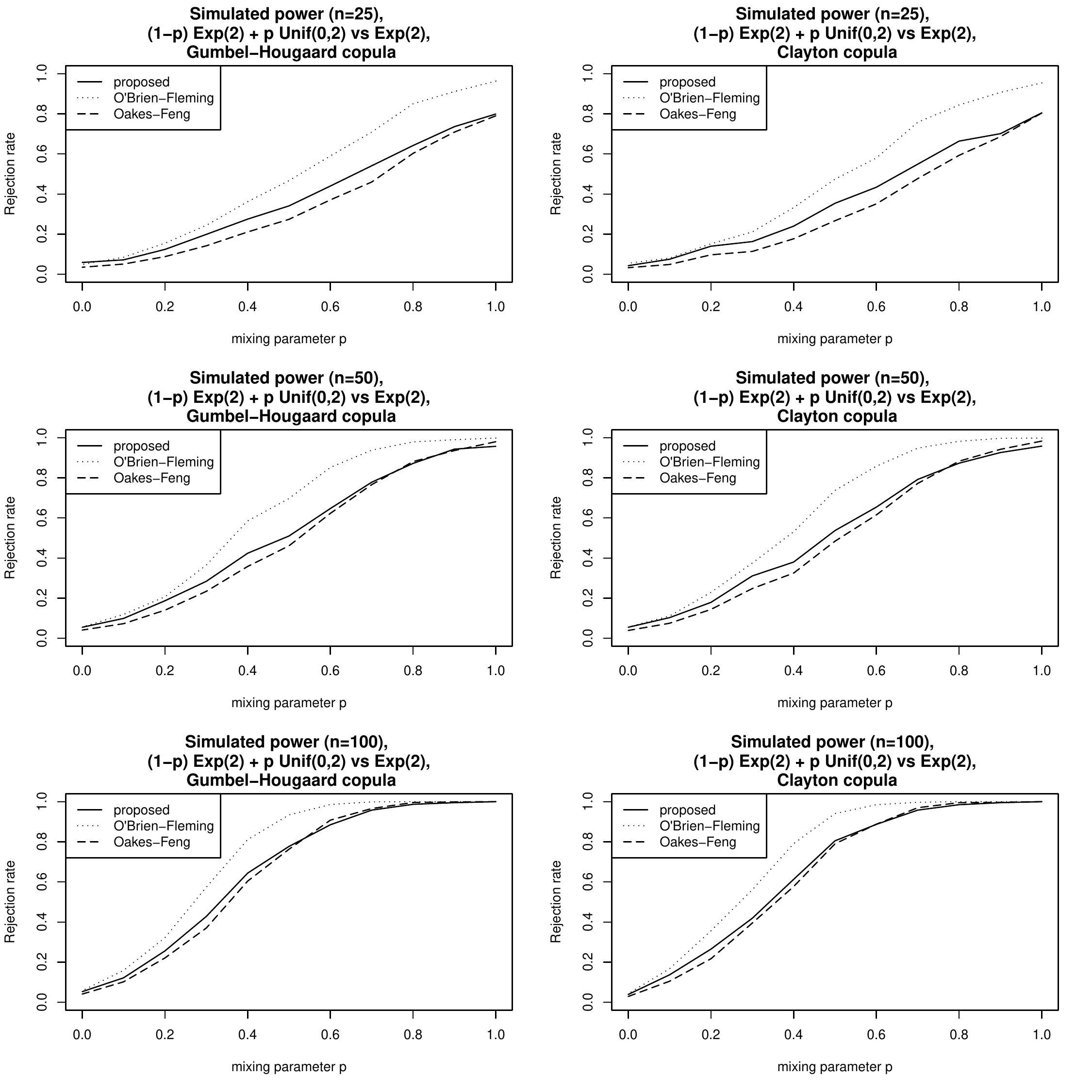}
	\caption{Simulated power of three selected right-tailed randomization-based tests with significance level $\alpha=5\%$.}
	\label{fig:power_mix1_exp}
\end{figure}

For now, we only focus on the results for Scenario~1 graphically presented in Figure~\ref{fig:power_mix1_exp};
the results for Scenarios~2 and~3 are presented in Section~\ref{app:more_power} of the Supplementary Material and they are similar to those presented here.
We can see from Figure~\ref{fig:power_mix1_exp} by comparing each combination of left and right panel that the copula that connects the lifetimes apparently has only little influence on the performance of the tests.
Not surprisingly, the power of all tests increases when the sample size increases (top to bottom in the figure) and when we depart from the null hypothesis (from left to right within each panel).
The paired Prentice-Wilcoxon test\cite{brien87} is always the most powerful one.
In most cases, the proposed test has the next higher power but its performance is generally very similar to that of the stratified log-rank test\cite{oakes10}.

Multiple comments are in order.
First, in this simulation study we could confirm the earlier findings\cite{woolson92} that the paired Prentice-Wilcoxon test is indeed quite powerful. This is also not surprising because it is based on an efficient score approach.
Next, the power of the stratified log-rank test could possibly be greater if the optimal combination with the unstratified log-rank test was used.\cite{oakes10}
However, an implementation of the combination would be beyond the scope of the present paper.
In addition, it should not be forgotten that the considered competitor tests were proposed for the sharp null hypothesis of equal survival distributions; it is only natural that such tests potentially have a greater power than the proposed test which was designed for the weak null hypothesis $H_0: \theta=0.5$.
Another reason for the relatively high power of the log-rank test and the paired Prentice-Wilcoxon test is the fact that the generated lifetimes were not truncated at $\tau$, whereas smaller values of $\tau$ let the relative treatment effect get closer to $0.5$, i.e., closer to the null hypothesis.
In the light of these facts, the power of the proposed test procedure is very competitive.
In addition, as initially mentioned, our focus was on the development of an easily interpretable estimand-based inference procedure, rather than developing the most powerful test for comparing two survival functions.

\section{Data example}
\label{sec:data}

We illustrate our methodology by re-analyzing a well known benchmark data set which has been published in the \texttt{R} package \texttt{survival}\cite{survival-package}. The data set \texttt{diabetic}, in detail described and analyzed by \citeauthor{huster89}\cite{huster89}, contains $394$ observations from a trial including $197$ patients with ``high-risk" diabetic retinopathy, a complication associated with diabetes mellitus that frequently leads to blindness.
In this trial each patient acts as its own control: one eye was randomized to a laser photocoagulation, while the other eye received no treatment. 

Apart from suffering from diabetic retinopathy, the inclusion criterion of the trial was a visual acuity of at least 20/100 in both eyes.
The aim of the study was to investigate the effect of the laser treatment on delaying the onset of blindness, defined by a visual acuity of less than 5/200 at two consecutive visits after four months. 
Thus, the survival times are the times until blindness for the eyes. Censoring was caused by death, dropout, or the end of the study.
Consequently, we consider the censored paired survival outcomes and their relative treatment effect.

The data set consists of two subgroups, defined by the type of diabetes, i.e., patients with juvenile onset diabetes (diagnosis before an age of 20, 114 patients) and adult onset diabetes (83 patients).
%In the latter subgroup, $85.1\%$ of the patients had at least one censoring, in the juvenile group it were $78.9\%$. 
In these subgroups, respectively $78.9\%$ and $85.1\%$ of the patients had at least one censoring.
Further, there are several other covariates, e.g., the laser type and a risk score.

\begin{figure}[ht]
	\centering
	%\mbox{
		%\phantom{X} \hspace{-1.2cm}
		\begin{subfigure}{1\textwidth}
			\centering
			\includegraphics[width=0.55\textwidth]{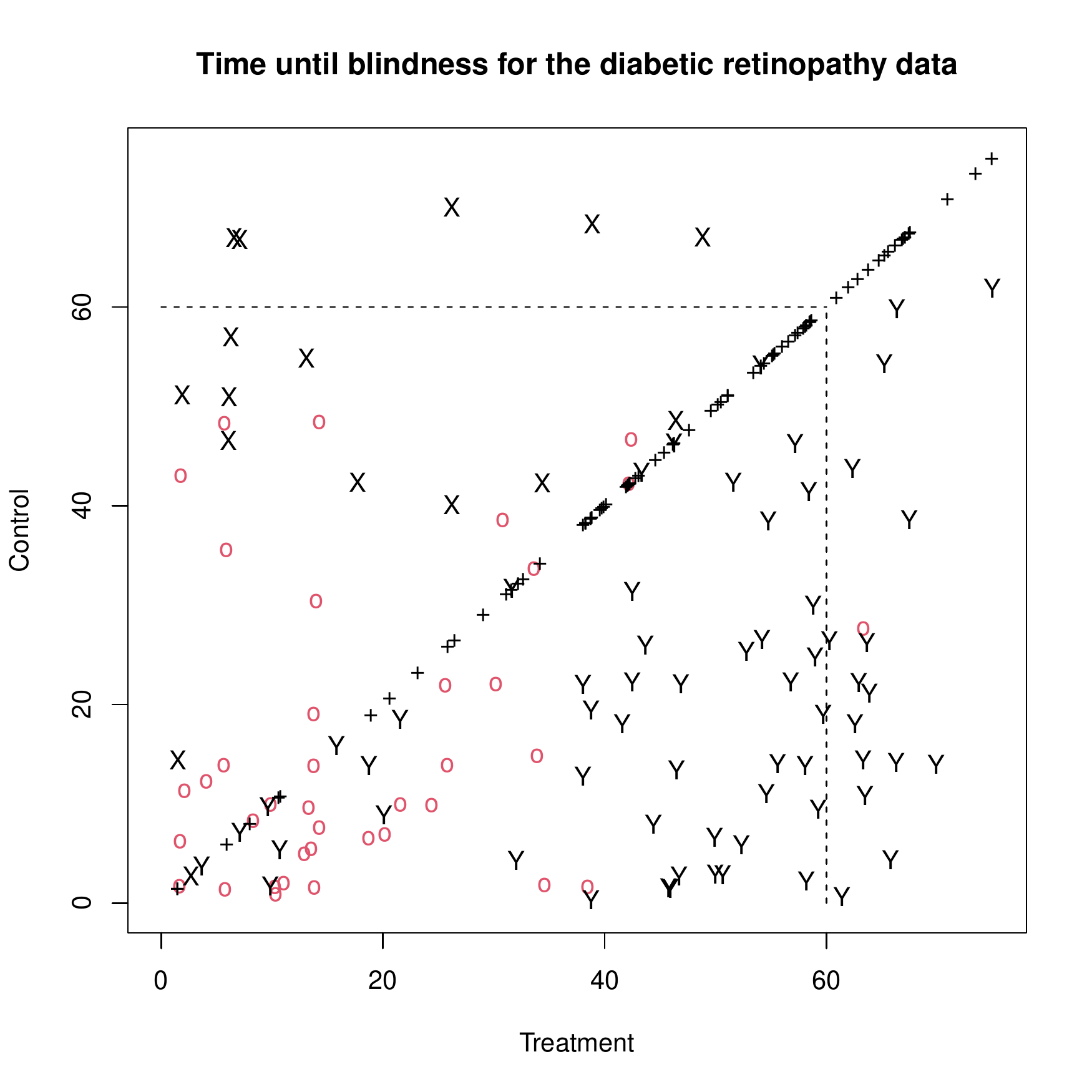}\\
			(a) Time until blindness for the diabetic retinopathy data. The values on the diagonal (denoted by "+") correspond to patients where no blindness occurred throughout the observational period. Red circles indicate blindness of both eyes, values below the diagonal (denoted by "Y") indicate blindness of the control eye only, whereas values above the diagonal (denoted by "X") indicate blindness of the treated eye only.
			\label{fig:kaplan_meier_plot_a}
		\end{subfigure}
		\\[0.3cm]
		\begin{subfigure}{1\textwidth}
			%\phantom{X} \hspace{-1.2cm}
			\centering
			\includegraphics[width=0.55\textwidth]{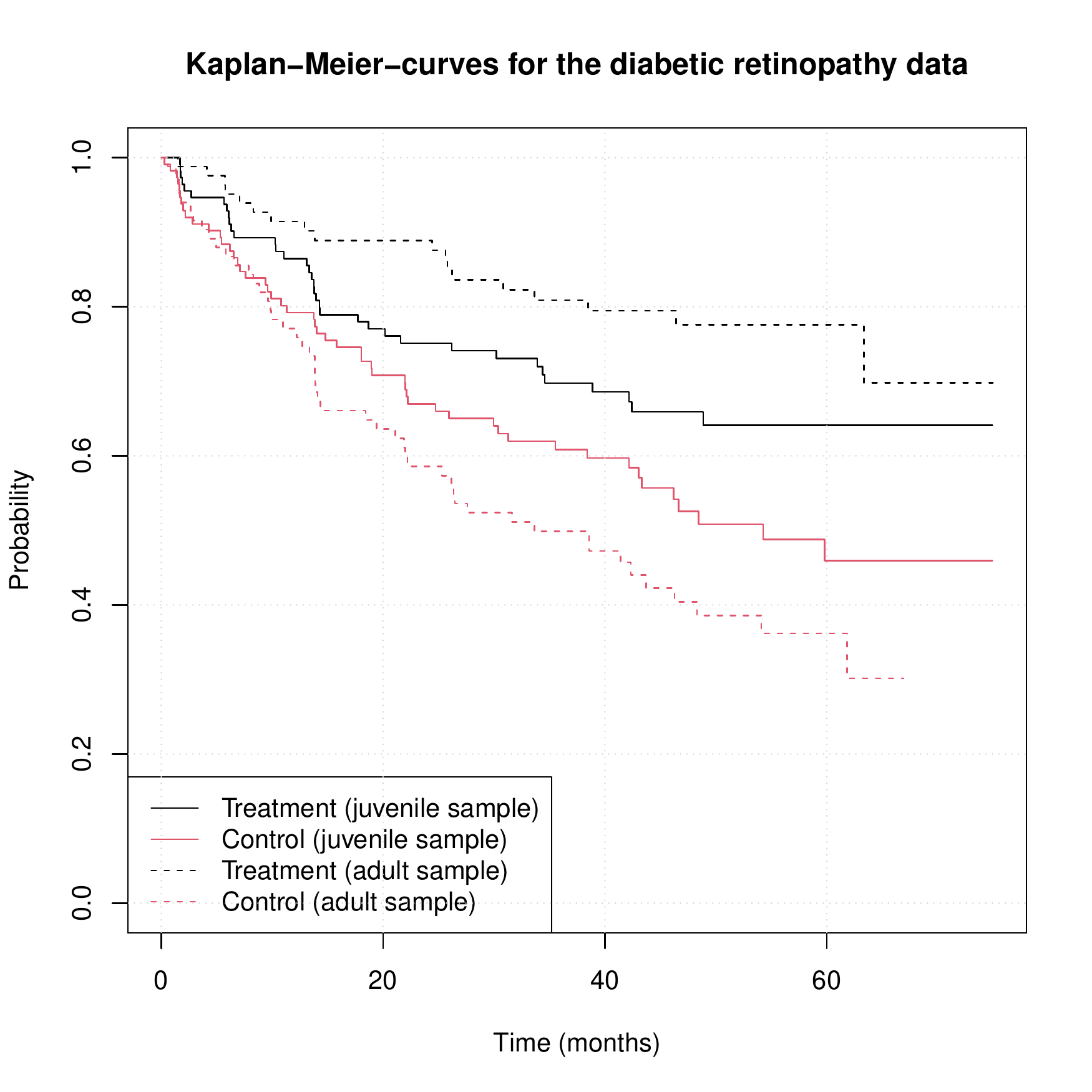}\\
			(b) Kaplan-Meier-curves for the two eyes (treated and control) for the diabetic retinopathy data, fitted separately for the juvenile and the adult sample.
			\label{fig:kaplan_meier_plot_b}
			%\phantom{X}\\\vspace{0.8cm}\phantom{X}
		\end{subfigure}
		%}
	\caption{Graphical summaries of the data set.}
	\label{fig:data}
\end{figure}

For a better overview of the data, Figure \ref{fig:data}(a) visualizes the time until blindness for the total sample, regardless of the age at diagnosis. From the pattern of the observations displayed in the figure, we conclude that the laser treatment seems in general to delay blindness.
For a separate investigation of the juvenile and adult sample, respectively, Figure \ref{fig:data}(b) displays the estimated Kaplan-Meier curves for each eye, i.e. the one achieving a laser treatment and the other one acting as the control. For both subsamples, we observe a visible difference between the Kaplan-Meier curves and again we conclude that the laser treatment seems to delay the onset of blindness compared to the control, which becomes even more visible in the adult sample.  
%Further, Figure \ref{fig:kaplan_meier_plot}(b) indicates that, regardless of the treatment, there don't seem to be differences related to the age at diagnosis only.

$$  \text{Figure~\ref{fig:data} about here.} $$

$$  \text{Table~\ref{tab:case_study} about here.} $$

\begin{table}[ht]
	\caption{Comparison of $95\%$- confidence intervals and p-values for the diabetic retinopathy data. Abbreviations: asy.\ = asymptotical normal, bs.\ = bootstrap, rand.\ = randomization; lin.\ = linear, tra.\ = $\log$-$\log$-transformed}
		\centering
	\vspace{0.2cm}
	\begin{tabular}{|cc|cc|cc|}
\hline\noalign{\smallskip}
		\multicolumn{2}{|c|}{\multirow{2}{*}{Method}} &\multicolumn{2}{c|}{ Juvenile sample}&\multicolumn{2}{c|}{Adult sample}\tabularnewline
		& & $95\%$- confidence interval & p-value &$95\%$- confidence interval & p-value  \tabularnewline
		\hline
		\multirow{2}{*}{asy.}& lin. & \big[0.517, 0.678\big]& 0.017 & \big[0.655, 0.807\big] & $<0.001$ \tabularnewline
		& tra. & \big[0.513, 0.673\big]& 0.025 & \big[0.646, 0.798\big] & $<0.001$ \tabularnewline
		
		\multirow{2}{*}{bs.}& lin. & \big[0.514, 0.680\big]& 0.014 & \big[0.652, 0.802\big]& $<0.001$\tabularnewline
		&tra. & \big[0.517, 0.677\big]& 0.012 & \big[0.655, 0.800\big]& $<0.001$\tabularnewline
		
		\multirow{2}{*}{rand.}& lin. & \big[0.515, 0.680\big]& 0.025 & \big[0.654, 0.809\big]& $<0.001$\tabularnewline
		&tra. & \big[0.515, 0.676\big]& 0.025 & \big[0.651, 0.801\big]& $<0.001$\tabularnewline

		\hline
	\end{tabular}
	\label{tab:case_study}
\end{table}

In order to confirm our visual findings, we will estimate the relative treatment effect of the laser photocoagulation and the corresponding confidence intervals, as well as  performing the corresponding two-sided hypothesis tests in order to assess the effectiveness of this therapy. 
We fix the maximum follow-up time as $\tau=60$ (indicated by the dashed box in Figure \ref{fig:data}(a)) and perform the analysis for both samples, that is juvenile and adult onset diabetes, separately. For the juvenile sample, we obtain a relative treatment effect of $\hat\theta_J=0.598$, for the adult sample we have $\hat\theta_A=0.731$. The corresponding $95\%-$ confidence intervals and the results of the (two-sided) hypothesis test for the different approaches described in Section \ref{sec:inference} are summarized in Table~\ref{tab:case_study}. The transformation used for the analysis is given by $\phi(\theta)=\log(-\log(\theta))$ and all results based on bootstrap were achieved by using $B=$2,000 bootstrap repetitions. 
For each subsample, all confidence intervals and also all p-values are very similar.
%Regardless of the sample under consideration, we observe that the asymptotic confidence bands are wider than the ones based on bootstrap and re-randomization, respectively, where the latter two are very similar. Consequently, this results in a higher p-value of the corresponding asymptotic tests compared to the other test results. 
In general, the effect of an additional transformation of the test statistic is rather small and yields very similar confidence intervals and test results.
As already indicated by Figure \ref{fig:data}(b), there is a notable difference of treatment and control eye, which is even larger in the adult sample. This is confirmed by p-values  below $0.001$ for the adult sample for all tests under consideration. For the juvenile sample, p-values lie between $0.012$ and $0.025$. Hence, we conclude a significant treatment effect for both samples at the significance level $\alpha=5\%$.

Our results are similar to the findings from \citeauthor{oakes10}\cite{oakes10}, who investigated the same data set regarding the treatment effect. In their paper, they propose three different test approaches and conclude that, for the example at hand, all resulting treatment effects are significant.

%diabetic retinopathy (survival package)
%\\
%covariates: age, risk score. (possibly for factorial analysis)
%Similar $p$-value as in the Oakes and Feng paper!

%Other possible data sets:
%\noindent rats (survival package): three rats per litter; two, say $T_1,T_2$ treated with standard treatment, another one, say $T_3$, treated with new(?) treatment.
%\\
%Analysis with above-described method should pose no problem, simply take
%$$ \hat p = \frac12(\hat p_1 + \hat p_2) = \frac12 (\hat P(T_1 > T_3) + \hat P(T_2 > T_3)) + \frac14( \hat P(T_1 =T_3) + \hat P(T_2 = T_3)). $$
%(Possibly apply permutation iterations for randomizing $T_1,T_2$, so as to minimize the influence of the specific groupings.)
%\\
%As of yet, it is unknown how to use the data in a more sophisticated way.

\section{Discussion}
\label{sec:disc}

%In this paper, we have developed a competing risks-based hypothesis test for the superiority of one treatment over another in the context of matched pairs studies.
%One enormous benefit of the proposed test is that it is based on an estimator for the relative treatment effect which is easily interpretable in terms of survival times, unlike log-rank or score-based tests.
%Yet, our test still exhibited in simulations a good power behaviour under alternatives.
%Additionally, the test gives rise to confidence intervals for the relative treatment effect.

%From the power simulations, we could also see that the dependence structure of the lifetimes apparently plays an almost negligible role.
%The very similar $p$-values obtained from the test based on a, say, marginal variant of the relative treatment effect which acknowledges the possible correlation only in the studentization \citep{dobler22} seems to support the little importance of the copula structure.
%However, our simulations regarding the type-I error control let us be aware that, for some unfavourable distributional combinations, quite high sample sizes might be needed to keep the significance level, even for the randomization test variant. 
%Here, the underlying copula seems to be important.

In this paper, we developed a new estimand in the context of paired, right-censored survival data, the so-called relative treatment effect, to compare the effectiveness of two treatments. 
Such data occur for instance in matched pairs studies.
We derived confidence intervals and hypothesis tests and could demonstrate all desirable propoerties by means of a simulation study.

The relative treatment effect $\theta$ quantifies the stochastic ordering of treatment outcomes but not how much bigger one survival times is than the other.
	In this sense, it is a global measure for the superiority of the first treatment, although other measures, e.g., about the actual size of the differences, might also be of major importance in some applications.
	One extension of the present method could be the incorporation of additional patient covariates, e.g., those that are used for the matching of individuals into pairs, which could be used to tackle the classification problem of who specifically should receive which treatment.
	This could be achieved in terms of semiparametric regression models or by involving relative treatment effects in a machine learning algorithm.

There are multiple possibilities for other extensions of the present approach.
One open question is how to incorporate additional patients that could not be matched with others or if multiple patients of one treatment group could be matched with just one patient of the other group.
The latter problem could potentially be approached by means of an appropriate re-weighting of the within-pair comparisons.
However, there is the risk that the celebrated easy interpretability of the relative treatment effect could be lost.
Another way to extend the present approach is the incorporation of additional covariates.
Due to the favourable competing risks approach, such an extension could be achieved rather straightforwardly, e.g., by means of cause-specific hazard models or subdistribution hazard models.

Finally, we note that the hypothesis test presented in this paper investigates the significance of the treatment effect. However, there might also occur situations where one is rather interested in testing whether the deviation of the treatment effect of $0.5$ is not larger than pre-specified values $\epsilon_1$ and $\epsilon_2$, respectively. 
In other words, this requires an equivalence test for $H_0^{eq}: \theta \notin (0.5 - \varepsilon_1, 0.5 + \varepsilon_2)$ against $H_a^{eq}: \theta \in (0.5 - \varepsilon_1, 0.5 + \varepsilon_2)$
%Precisely this means testing $H_0^{eq}: \theta \notin (0.5 - \varepsilon_1, 0.5 + \varepsilon_2)$ against $H_a^{eq}: \theta \in (0.5 - \varepsilon_1, 0.5 + \varepsilon_2)$, constituting an equivalence test 
(see, for example, \citeauthor{wellek2002}\citealp{wellek2002}). Such an approach could provide a very flexible framework for statistical inference, address numerous other research questions, and consequently provide a useful addition to the test proposed in this paper. We leave the development of such a procedure for future research. 

\section*{Acknowledgements}

The authors wish to thank Arthur Allignol and Mark Clements for providing a suitable update of the \emph{etm} R package, Marialuisa Restaino, % for pointing out some references in the field of bivariate survival distribution estimation, and 
Nan van Geloven, Liesbeth de Wreede, Hein Putter, Giuliana Cortese, Thomas Scheike, Katharina Kramer, and Paul Blanche for useful discussions.

%\section*{Supplementary Material}

%Supplementary Material containing technical details about the competing risks-based estimation approach, all proofs, and additional simulation results is available online.
 
%
\section*{Conflict of interest}
The authors declare that they have no conflict of interest.

\section*{Data Availability Statement}

The R code used for the simulation studies, for generating synthetical data sets, and for the real data application is available at \url{https://github.com/dennis-dobler/relative\_treatment\_effect\_paired\_survival}. 
%
%\newpage
%\phantom{X}
%\newpage
%\phantom{X}
%\newpage

%\bibliography{literature.bib}

%\section*{Bibliography}

\appendix

\section{Technical details for the competing risks-based estimation of the relative treatment effect}
\label{supp:cr}

In this section, we will explain in detail how the transformation of the paired, right-censored data to competing risks data can be achieved and why these competing risks data are usable for valid estimation of the relative treatment effect.

Let us thus assume that a data set consists of independently and identically distributed data points $(X_{i1}, \delta_{i1
}, X_{i2}, \delta_{i2}), i=1,\dots, n $, as described in Section~\ref{sec:RTE} in the main manuscript.
As explained there, we wish to estimate the relative treatment effect $\theta$ with the help of a competing risks approach.
Our key strategy is to transform the data into a  competing risks data set: $$(Z_i, \varepsilon_i) = (\min(\check T_i, \check C_i), \check \varepsilon_i \cdot  1\{\check T_i \leq  \check C_i\}), \quad i=1,\dots, n,$$
where $\check T_i=\min(T_{i1}, T_{i2}, \tau)$ with survival function $S(t)=P(\check T_i > t) \ (0\leq t \leq \tau)$ denote the event times, $\check C_i = \min(C_{i1}, C_{i2})$ with survival function $G(t)=P(\check C_i > t) \ (0\leq t \leq \tau)$ denote the censoring times, and $\check\varepsilon_i \in \{1,2,3\}$ are the event indicators, respectively.
Again, $\check T_i, \check C_i$, and $\check \varepsilon_i$ are not fully observable but $(Z_i, \varepsilon_i)$ is.
The three different (artificial) events are defined as follows:
\begin{itemize}
	\item[1.] If the first treated subject experienced an \emph{observable} event before the second, i.e., $X_{i1} < X_{i2}, \delta_{i1} = 1$  or if $X_{i1} = X_{i2}, (\delta_{i1}, \delta_{i2}) = (1,0)$, we say the first type of event ($\varepsilon_i =1$) took place at time $X_{i1}=T_{i1}$.
	Note though that the second just mentioned case is impossible due to the independent and continuous censoring assumptions:
	%we have for the letter case
	$$P(X_{i1}=X_{i2}, \delta_{i1}=1, \delta_{i2}=0)  \leq P(T_{i1} = C_{i2} ) = \int P(C_{i2} = t) d P^{T_{i1}}(t) = 0.$$
	%
	%$P(X_{1i} = X_{2i},\delta_{1i}=1,  \delta_{2i}=0) \leq P(C_{2i}=t) =0 $.
	\item[2.] Similarly, the second type of event ($\varepsilon_i =2$) took place at time $X_{i2}=T_{i2}$ if  $X_{i2} < X_{i1}$ and $\delta_{i2} = 1$.
	%(or if $X_{1i} = X_{2i}$ and $(\delta_{1i}, \delta_{2i}) = (0,1)$).
	\item[3.] If both events were \emph{observed} simultaneously, i.e.,  $X_{i1} = X_{i2}$ and $\delta_{i1} = \delta_{i2} = 1$, the third type of event ($\varepsilon_i =3$) took place at $X_{i1}= X_{i2}=T_{i1}=T_{i2}$.
\end{itemize}
In all other cases, the observation is censored ($\varepsilon_i=0$) at $$\min(X_{i1}, X_{i2})=\min(C_{i1},C_{i2}).$$

In the following, we suppress the index $i$ for ease of presentation when there is no need to specify the precise subject. 
We denote the cause-specific cumulative hazard functions in the competing risks framework by $A_j, j=1,2,3$. 
The following lemma is the crucial step towards estimating the relative treatment effect $\theta$; in fact, it shows that no relevant information is lost by the above-described conversion to a competing risks data set.

\begin{lemma}
	\label{lem:nae}
	The cause-specific Nelson-Aalen estimators 
	$$\widehat A_{j,n}(t) = \sum_{u \leq t} \tfrac{\sum_{i=1}^n 1\{ Z_i = u , \varepsilon_i = j\}}{\sum_{i=1}^n1\{ Z_i \geq u \}}$$ 
	are uniformly consistent for $A_j(t), j=1,2,3$ in $t \in [0,\tau]$ as $n\to\infty$.
	In addition, $\{n^{1/2} (\widehat A_{j,n} - A_{j})\}_{j=1}^3$ converges in distribution to a three-dimensional zero-mean Gaussian process as $n\to\infty$.
\end{lemma}

The proof of Lemma~\ref{lem:nae} (provided in Section~\ref{supp:proofs} below) reveals that the transformation of the paired survival data into competing risks data preserves the underlying intensities. As a consequence,
other properties of the Nelson-Aalen estimators are retrieved, such as their interpretation as nonparametric maximum likelihood estimators.
Another consequence of Lemma~\ref{lem:nae} is that the Aalen-Johansen estimators\citep{aalen78} for the cumulative incidence functions for all event times are similarly estimable; cf.\ \citeauthor{dobler17}\cite{dobler17} for the general case with both continuous and discrete components in the event time distribution.
To be specific, estimation of the relative treatment effect is achievable as follows:
$$ \widehat \theta_n = \widehat F_{2,n}(\tau) + \tfrac12 \widehat F_{3,n}(\tau) = \int_0^\tau \widehat S_n(u-) d (\widehat A_{2,n} + \tfrac12 \widehat A_{3,n})(u), $$
where $\widehat S_n(t) = \prod_{u \leq t} \{1 - d(\widehat A_{1,n} + \widehat A_{2,n} + \widehat A_{3,n})(u)\}$ denotes the Kaplan-Meier estimator of $S(t)= P(\check T > t)$ and $\widehat F_{j,n}$ are the Aalen-Johansen estimators of $F_j$, $j=2,3$.
The minus sign in an argument indicates the left-continuous version of a function.

\section{Asymptotic variances of the relative treatment effect estimators, and consistent variance estimators}
\label{supp:var}

\subsection{Asymptotic variances of $\sqrt{n}(\widehat \theta_n - \theta)$ and $\sqrt{n}(\tilde \theta_n - \tfrac12)$}

{Tedious but straightforward calculations revealed the following expression for the asymptotic variance of the relative treatment effect estimator, $\sqrt{n}(\widehat \theta_n - \theta)$:}
\begin{align*}
	\sigma^2_\theta &= \int_0^\tau \int_0^\tau  S(u-)S(v-)  \Big[ \int_0^{\min(u,v)-} \frac{\sigma^2_\bullet(dw)}{1-\Delta A_\bullet(w)}  (A_2 + \tfrac12 A_3)(du) (A_2 + \tfrac12 A_3)(dv) \\
	& - 2 \int_0^{\min(u,v)-} \frac{(\sigma_{12} + \tfrac12\sigma_{13} + \sigma_2^2 + \tfrac32 \sigma_{23} +\tfrac12 \sigma_3^2)(dw)}{1-\Delta A_\bullet(w)}  (A_2 + \tfrac12 A_3)(du) A_\bullet(dv) \\
	& + (\sigma_2^2 + \sigma_{23} + \tfrac14 \sigma_3^2)(\min(u,v)) A_\bullet(d u) A_\bullet(dv)\Big],
\end{align*}
where $A_\bullet = A_1+A_2+A_3$ is the all-cause cumulative hazard function,  
$$\sigma_j^2(t) = \int_0^t \frac{1-\Delta A_j(u)}{S(u-)G(u-)} A_j(du), \ j=1,2,3,$$ and
$$\sigma_{j\ell}(t) = - \sum_{u \leq t} \frac{\Delta A_j(u) \Delta A_\ell(u)}{S(u-)G(u-)}, \ j\neq \ell,$$ are the asymptotic variance and covariance functions of the normalized cause-specific Nelson-Aalen estimators, respectively,
and $$\sigma_\bullet^2(t) = \int_0^t \frac{1-\Delta A_\bullet(u)}{S(u-)G(u-)} A_\bullet(du) = \sum_{j\neq \ell} (\sigma^2_j(t) + \sigma_{j\ell}(t)), \quad t \in [0,\tau],$$ is the asymptotic variance function of the normalized all-cause Nelson-Aalen estimator.
{We propose to use the consistent Greenwood-type variance and estimators of $\sigma_j^2$ and $\sigma_{j\ell}$; see, e.g., formulas (4.4.17) and (4.4.18) in \citeauthor{abgk93}\cite{abgk93}.}

{For the randomization version of the relative treatment effect estimator, $\sqrt{n}(\tilde \theta_n - \tfrac12)$, we discovered a structure similar to $\sigma^2_\theta$, except that all quantities $ \sigma_j^2, \sigma_{j\ell}, A_j $ are replaced by their randomization counterparts; see the subsequent subsection for details.}
%i.e.\ the limits of, say, $ \tilde{ \widehat \sigma}_j^2,  \tilde{ \widehat\sigma}_{j\ell},  \tilde{ \widehat A}_j $ in probability, $j,\ell = 1,2,3, \ j \neq \ell$.}  

\subsection{Variance estimators}

{As motivated in the previous subsection, a consistent estimator for the variance of the normalized randomized relative treatment effect, $\sqrt{n}(\tilde \theta_n - \tfrac12)$,} is given by
% \begin{align*}
%  \tilde\sigma^2_{\theta,n} & = \int_0^\tau \int_0^\tau  \tilde S_n(u-)\tilde S_n(v-)  \Big[ \int_0^{\min(u,v)-} \frac{\tilde \sigma^2_{\bullet,n}(dw)}{1-\Delta \tilde A_{\bullet,n}(w)}  (\tilde A_{2,n} + \tfrac12 \tilde A_{3,n})(du) (\tilde A_{2,n} + \tfrac12 \tilde A_{3,n})(dv) \\
% & - 2 \int_0^{\min(u,v)-} \frac{(\tilde \sigma_{12,n} + \tfrac12\tilde \sigma_{13,n} + \tilde \sigma_{2,n}^2 + \tfrac32 \tilde \sigma_{23,n} +\tfrac12 \tilde \sigma_{3,n}^2)(dw)}{1-\Delta \tilde A_{\bullet,n}(w)}  (\tilde A_{2,n} + \tfrac12 \tilde A_{3,n})(du) \tilde A_{\bullet,n}(dv) \\
% & + (\tilde \sigma_{2,n}^2 + \tilde \sigma_{23,n} + \tfrac14 \tilde \sigma_{3,n}^2)(\min(u,v)) \tilde A_{\bullet,n}(d u) \tilde A_{\bullet,n}(dv)\Big]
% \end{align*}
\begin{align*}
&	\tilde\sigma^2_{\theta,n}  = \int_0^\tau \int_0^\tau  \tilde S_n(u-)\tilde S_n(v-)  \Big[ \int_0^{\min(u,v)-} \frac{\tilde \sigma^2_{\bullet,n}(dw)}{1-\Delta \tilde A_{\bullet,n}(w)}  (\tilde A_{2,n} + \tfrac12 \tilde A_{3,n})(du)\\
&\cdot (\tilde A_{2,n} + \tfrac12 \tilde A_{3,n})(dv) - 2 \int_0^{\min(u,v)-} \frac{(\tilde \sigma_{12,n} + \tfrac12\tilde \sigma_{13,n} + \tilde \sigma_{2,n}^2 + \tfrac32 \tilde \sigma_{23,n} +\tfrac12 \tilde \sigma_{3,n}^2)(dw)}{1-\Delta \tilde A_{\bullet,n}(w)} \\
&\cdot (\tilde A_{2,n} + \tfrac12 \tilde A_{3,n})(du) \tilde A_{\bullet,n}(dv) 
+ (\tilde \sigma_{2,n}^2 + \tilde \sigma_{23,n} + \tfrac14 \tilde \sigma_{3,n}^2)(\min(u,v)) \tilde A_{\bullet,n}(d u) \tilde A_{\bullet,n}(dv)\Big]
\end{align*}
with $\tilde S_n$ and $\tilde A_{j,n}$ being the Kaplan-Meier and the cause-specific Nelson-Aalen estimators, respectively, based on the randomized sample $(\tilde Z_i, \tilde \varepsilon_i), i=1, \dots, n$, and similarly for the estimators $\tilde \sigma_{j\ell,n}$ and $\tilde \sigma^2_{j,n}$ of $\sigma_{j\ell}$ and $\sigma^2_{j}$, respectively, $j=1,2,3$, $\ell \neq j$.
The estimator $\widehat \sigma_{\theta,n}^2$ {of $ \sigma_{\theta}^2 $} is similarly obtained, just based on the original sample $( Z_i, \varepsilon_i), i=1, \dots, n$.

\section{Proofs}
\label{supp:proofs}
%\subsection*{Proofs}

\textit{Proof of Lemma~\ref{lem:nae}.}
%\begin{proof}[Proof of Lemma~\ref{lem:nae}]
We are going to show that the censored competing risks data set exhibits the correct underlying hazard rates. The consistency and asymptotic normality statements then follow from well-known results in the literature, e.g., \citeauthor{abgk93}\cite{abgk93}. 
We refer to \citeauthor{dobler17}\cite{dobler17} for detailed derivations regarding the general case of event times with both discrete and continuous components.

We first consider continuous components of the hazard rate.
Thus, at points of continuity $t < \tau$ of the distribution of $\check T$, we have 
\begin{align*}
	& \lim_{dt \downarrow 0} P(Z \in [t, t + dt], \varepsilon=1 \ |  \ Z \geq t) / dt
	\\
	& = \lim_{dt \downarrow 0} P( \{ X_1 < X_2, \delta_1 = 1, X_1 \in [t, t+dt] \} \\
	&\cup \{ X_1 = X_2 \in [t,t+dt], \delta_1=1, \delta_2=0 \} 
	| \ X_1 \geq t, X_2 \geq t) / dt \\
	& =\lim_{dt \downarrow 0} \frac{P(T_1 \in [t, t + dt], T_2 \geq t + d t, C_1 \geq t + d t,  C_2 \geq t  ) }{P( T_1 \geq t, T_2 \geq t, C_1 \geq t, C_2 \geq t)  d t} \\
	& =\lim_{dt \downarrow 0} \frac{P(T_1 \in [t, t + dt], T_2 \geq t + d t) P(C_1 \geq t + dt, C_2 \geq t)}{P( T_1 \geq t, T_2 \geq t) P(C_1 \geq t, C_2 \geq t) d t} \\
	& =\lim_{dt \downarrow 0} P(\check T \in [t,t+dt], \check \varepsilon = 1 \ | \  \check T \geq t) / dt = \lim_{dt \downarrow 0}  d A_1(t)/dt.
\end{align*}

A similar derivation holds for time points $t \leq \tau$ of discontinuity:
\begin{align*}
	& P(Z = t, \varepsilon=1 \ |  \ Z \geq t) 
	\\
	& =  P( X_1 = t,  X_2 > t, \delta_1 = 1 \ | \ X_1 \geq t, X_2 \geq t)  \\
	& = \frac{P(\min(T_1,\tau) = t, T_2 > t, C_1 \geq t, C_2 > t)}{P( T_1 \geq t, T_2 \geq t, C_1 \geq t, C_2 \geq t)} \\
	& = P(\check T =t , \check \varepsilon = 1 \ | \  \check T \geq t) = A_1(t) - A_1(t-).
\end{align*}
Here, $A_1(t-) = \lim_{u \uparrow t} A_1(u)$ denotes the left-continuous version of $A_1$.
The very same arguments can be used for the second competing risk, i.e., $\check\varepsilon = 2$.

For the third competing risk, we similarly have for any point $t < \tau$ of continuity that
\begin{align*}
	& \lim_{dt \downarrow 0} P(Z \in [t, t + dt], \varepsilon=3 \ |  \ Z \geq t) / dt
	\\
	& = \lim_{dt \downarrow 0} P( X_1 = X_2 \in [t, t+dt], \delta_1 = \delta_2 = 1 \ | \ X_1 \geq t, X_2 \geq t) / dt \\
	& = \lim_{dt \downarrow 0} \frac{P( T_1 = T_2 \in [t, t+dt], C_1 \geq t+dt, C_2 \geq t + dt)}
	{P(T_1 \geq t, T_2 \geq 2, C_1 \geq t, C_2 \geq t)  dt} \\
	& = \lim_{dt \downarrow 0} \frac{P( T_1 = T_2 \in [t, t+dt])P( C_1 \geq t+dt, C_2 \geq t + dt)}
	{P(T_1 \geq t, T_2 \geq t)P( C_1 \geq t, C_2 \geq t)  dt} \\
	& = \lim_{dt \downarrow 0} P(T_1 = T_2 \in [t,t+dt] \ | \ T \geq t) /dt \\
	& = \lim_{dt \downarrow 0}  P(\check T \in [t,t+dt] \ | \ \check T \geq t) /dt = \lim_{dt \downarrow 0} d A_3(t)/dt.
\end{align*}
Again, a similar derivation holds for times $t\leq \tau$ of discontinuity:
\begin{align*}
	& P(Z = t, \varepsilon=3 \ |  \ Z \geq t) 
	\\
	& = P( X_1 = X_2 = t, \delta_1 = \delta_2 = 1 \ | \ X_1 \geq t, X_2 \geq t)  \\
	& = \frac{P( \min(T_1,\tau) = \min(T_2,\tau) = t, C_1 \geq t, C_2 \geq t)}
	{P(T_1 \geq t, T_2 \geq 2, C_1 \geq t, C_2 \geq t) } \\
	& = \frac{P( \min(T_1,\tau) = \min(T_2,\tau) = t)P( C_1 \geq t, C_2 \geq t)}
	{P(T_1 \geq t, T_2 \geq t)P( C_1 \geq t, C_2 \geq t) } \\
	& = P(\min(T_1,\tau) = \min(T_2,\tau) =t \ | \ T_1  \geq t, T_2 \geq t) \\
	& =  P(\check T = t, \check \varepsilon =3 \ | \ \check T \geq t)  = A_3(t) - A_3(t-).
\end{align*}
This verifies that the Nelson-Aalen estimators estimate the correct quantities.
\hfill $\QED$
%\end{proof}
\\[0.2cm]
\noindent
\textit{Proof of Proposition~\ref{prop:1}.}
Part (a) is obvious.
\\[0.1cm]
(b) Lemma~\ref{lem:nae} revealed that the underlying intensity processes are preserved under the data transformation.
In addition, counting and at-risk process are known to be sufficient statistics for the nonparametric intensity processes\citep{aalen78b}.
It is thus apparent that the $\sigma$-algebra generated by the competing risks data set is sufficient for $\theta$.
\\[0.1cm]
(c) It is well-known that Aalen-Johansen estimators have an interpretation as NPMLEs; 
we refer to Sections~IV.4.1.5 and~IV.1.5 in \citeauthor{abgk93}\cite{abgk93} for a detailed discussion about the concept of NPMLEs, and	the NPMLE properties of the Aalen-Johansen and the more fundamental Nelson-Aalen estimators, respectively.
Likewise, it clear that $\widehat \theta_n$ is the NPMLE of $\theta$.
\hfill$\QED$
\\[0.2cm]
\noindent
\textit{Proof of Theorem~\ref{thm:main}.}
%\begin{proof}[Proof of Theorem~\ref{thm:main}]
The stated convergence in distribution is a consequence of the continuous mapping theorem in combination with the functional delta-method because the relative treatment effect estimator depends on the cause-specific Nelson-Aalen estimators through a combination of the following functionals: 
\begin{itemize}
	\item Wilcoxon functional (Section~3.9.4.1 in \citeauthor{vaart96}\citealp{vaart96}). For c\`adl\`ag functions $f$ and functions $g$ of bounded variation (bounded by a fixed constant), the functional $(f,g) \mapsto \int_0^\tau f(u) d g(u)$ is Hadamard-differentiable with derivative $(h, k) \mapsto \int_0^\tau h(u) dg(u) + \int_0^\tau f(u) d k(u)$, where the latter integral is defined via integration by parts: $\int_0^\tau f(u) d k(u) = f(\tau) k(\tau) - f(0) k(0) - \int_0^\tau k(u-) d f(u)$ if $k$ is not of bounded variation.
	\item Product integral (Section~3.9.4.5 in \citeauthor{vaart96}\citealp{vaart96}). For c\`adl\`ag functions $A$ of bounded variation (bounded by a fixed constant), the functional $A \mapsto \prod_{0 < t \leq \tau} (1 + A(dt))$ is Hadamard-differentiable with derivative 
	$$\alpha \mapsto \int_0^\tau \Big( \prod_{0 < s < t} (1 + A(ds)) \Big) \alpha(d t) \Big( \prod_{t < v \leq \tau} (1 + A(dv)) \Big) $$
	which is again defined by integration by parts.
	Here and below, $\prod$ with a time-continuous indexing stands for the product integral.
\end{itemize}

Due to the chain rule (Lemma~3.9.3 in \citeauthor{vaart96}\citealp{vaart96}), the composition of both functionals is also Hadamard-differentiable. 
Thus, since the Kaplan-Meier estimator satisfies $\widehat S_n(u) = \prod_{0 < s \leq u}(1 + (-\widehat A_{1,n}-\widehat A_{2,n} - \widehat A_{3,n})(du))$, we obtain the following asymptotic representation of the relative treatment effect estimator: 
\begin{align*}
	\sqrt{n} (\widehat \theta_n - \theta)  = & \int_0^\tau \sqrt{n} ( \widehat  S_n(u-) - S(u-) )  (A_2 + \tfrac12 A_3)(du) \\
	& + \int_0^\tau S(u-) \sqrt{n} \big[(\widehat A_{2,n} + \tfrac12 \widehat A_{3,n}) - (A_2 + \tfrac12 A_3) \big] (du) + o_p(1),
\end{align*}
where the first integral can be rewritten as 
\begin{align*}
	- \int_0^\tau S(u-) \int_0^{u-} \frac{\sqrt n  (\widehat A_{\bullet, n} -  A_\bullet)(dw)}{1 - \Delta A_\bullet (w)}  (A_2 + \tfrac12 A_3)(du) + o_p(1).
\end{align*}
Here, $\widehat A_{\bullet, n} = \sum_{j=1}^3\widehat A_{j, n}$ is the all-cause Nelson-Aalen estimator.
Simplifying the expression in the previous display once again, we get
\begin{align*}
	- \int_0^\tau S(u-) & \Big[ \sqrt n  (\widehat A_{\bullet, n} -  A_\bullet)(u-) 
	+ \sum_{w \in D_{u-}} \frac{\Delta A_\bullet(w)}{1-\Delta A_\bullet(w)} \cdot \Delta \sqrt n (\widehat A_{\bullet,n} - A_\bullet)(w)  \Big] \\
	& \times (A_2 + \tfrac12 A_3)(du) +
	o_p(1),
\end{align*}
where $D_{u-} \subset [0,u)$ denotes the set of all discontinuities of $A_\bullet$ in $[0,u)$.
In combination with the well-known limit behaviour of the Nelson-Aalen estimators on function spaces, 
our previous reasoning and the continuous mapping theorem establishes the asymptotic normality of the normalized relative treatment effect estimator.

It remains to verify the asymptotic variance $\sigma^2_\theta$. 
First note that, for deterministic c\`adl\`ag functions of bounded variation $a$ and $b$ and for stochastic processes $W$ and $Y$ that satisfy the required integrability conditions, Fubini's theorem implies that
$$ cov\Big(\int_0^\tau W(u) d a(u), \int_0^\tau Y (v) d b(v)\Big) = \int_0^\tau\int_0^\tau cov\big( W(u) , Y (v)\big) da(u) d b(v). $$
This, in combination with the asymptotic variances $\sigma^2_j$ and covariances $\sigma_{j\ell}$ of the normalized cause-specific Nelson-Aalen estimators
and straightforward but tedious computations reveal the claimed structure of asymptotic variance $\sigma^2_\theta$.
\hfill $\QED$
%\end{proof}
\\[0.2cm]
\noindent
\textit{Proof of Theorem~\ref{thm:rand}.}
%\begin{proof}[Proof of Theorem~\ref{thm:rand}]
At first, we verify the conditions of Theorem~2 in \citeauthor{dobler22}\cite{dobler22} to establish the asymptotic normality of the randomized relative treatment effect.
For this, we notice that the (randomized) Nelson-Aalen estimators $\tilde A_{j,n}$, $j=1,2,3$, are retrieved as functionals of the empiricial process of $(\tilde Z_i, \tilde \varepsilon_i), i=1, \dots, n$, indexed by the class
$$ \mathcal{F} = \{ (z,e) \mapsto 1\{ z \leq t, e=j \}, \  (z,e) \mapsto 1\{ z \geq t\}: \  0 \leq t \leq \tau; \  j=1,2,3  \}. $$
It is easy to see that $\mathcal F$ is a Vapnik-\u{C}ervonenkis class and it hence satisfies the uniform entropy condition (2.5.1) in \cite{vaart96}. 
As a consequence, $\mathcal F$ is both $\mathbb{P}$ and $\tilde{\mathbb{P}}$-Donsker and both distributions have bounded supremum norms, where $\mathbb{P}= P^{(Z,\varepsilon)}$ and $\tilde{\mathbb{P}} = P^{(\tilde Z,\tilde \varepsilon)}$.
Similarly, the class
$$ \tilde{\mathcal{F}} = \{ (z,e) \mapsto \tfrac12 [ f(z,j) + f(z,4-j)]: \ f \in \mathcal{F} \} $$
is $\mathbb{P}$-Donsker and $\mathbb P$ again has a bounded supremum norm with respect to $\tilde{\mathcal{F}}$.
Thus, Theorem~2 in \cite{dobler22} yields that the conditional distribution of $\sqrt{n}(\tilde \theta_n - 0.5)$ converges weakly to a normal distribution in probability as $n\to\infty$.

The limit variance, say $\tilde \sigma_\theta^2$, is in general different from $\sigma_\theta^2$.
That is why the studentization based on $\tilde \sigma_{\theta,n}^2$ is essential.
To justify the consistency of $\tilde \sigma_{\theta,n}^2$ for $\tilde \sigma_{\theta}^2$, note that this variance estimator is a continuous functional of the randomization empirical process, i.e., the empirical process based on $(\tilde Z_i, \tilde \varepsilon_i), i=1, \dots, n$.
Since the class $\mathcal F$ is Glivenko-Cantelli, the continuous mapping theorem yields the consistency of the variance estimator.
Finally, an application of a conditional version of Slutkzy's theorem concludes the proof.
\hfill$\QED$
%\end{proof}

%\newpage

%\appendixtwo
%\section*{Appendix 2}
\section{Additional simulation results}

\subsection{Sizes under $H_0: \theta=0.5$ and nominal significance levels $\alpha \in \{1\%, 10\%\}$}
\label{app:more_size}

Tables~\ref{tab:null_1} and~\ref{tab:null_10} summarize the simulation results for the size of the proposed one-sided tests under $H_0: \theta=0.5$ with nominal significance levels $\alpha=1\%$ and $10\%$, respectively.
The remaining simulation scenarios are the same as those which led to the results displayed in Table~\ref{tab:null_5} for $\alpha=5\%$ in the main manuscript.
As the overall impression is similar to the findings from that Table~\ref{tab:null_5}, no additional comments on the outcomes are given here.

\subsection{Additional power simulation results}
\label{app:more_power}

\begin{figure}[hb]
\centering
\includegraphics[width=\textwidth]{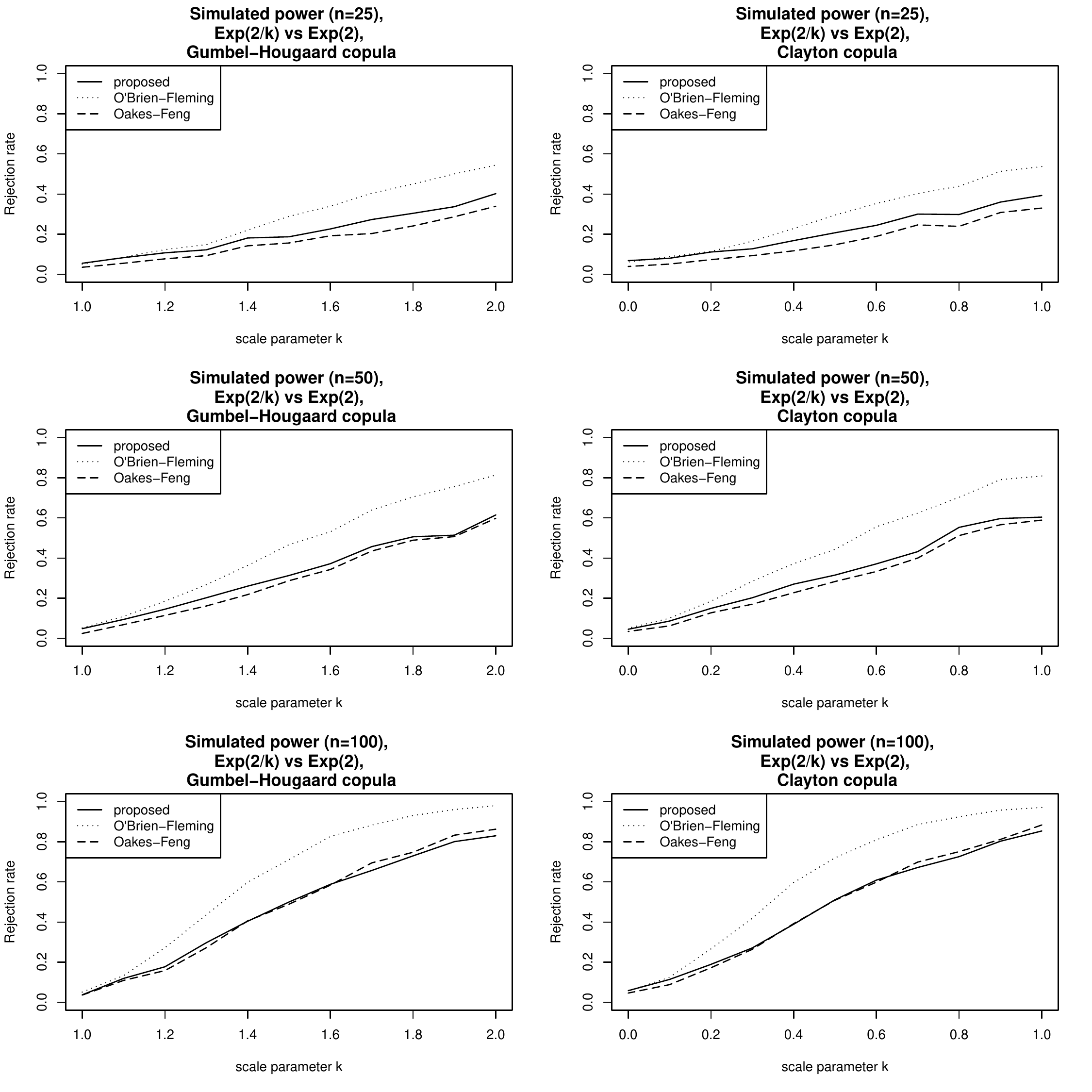}
\caption{Simulated power of three selected right-tailed randomization-based tests with significance level $\alpha=5\%$.}
\label{fig:power_exp_exp}
\end{figure}

\begin{figure}[ht]
\centering
\includegraphics[width=\textwidth]{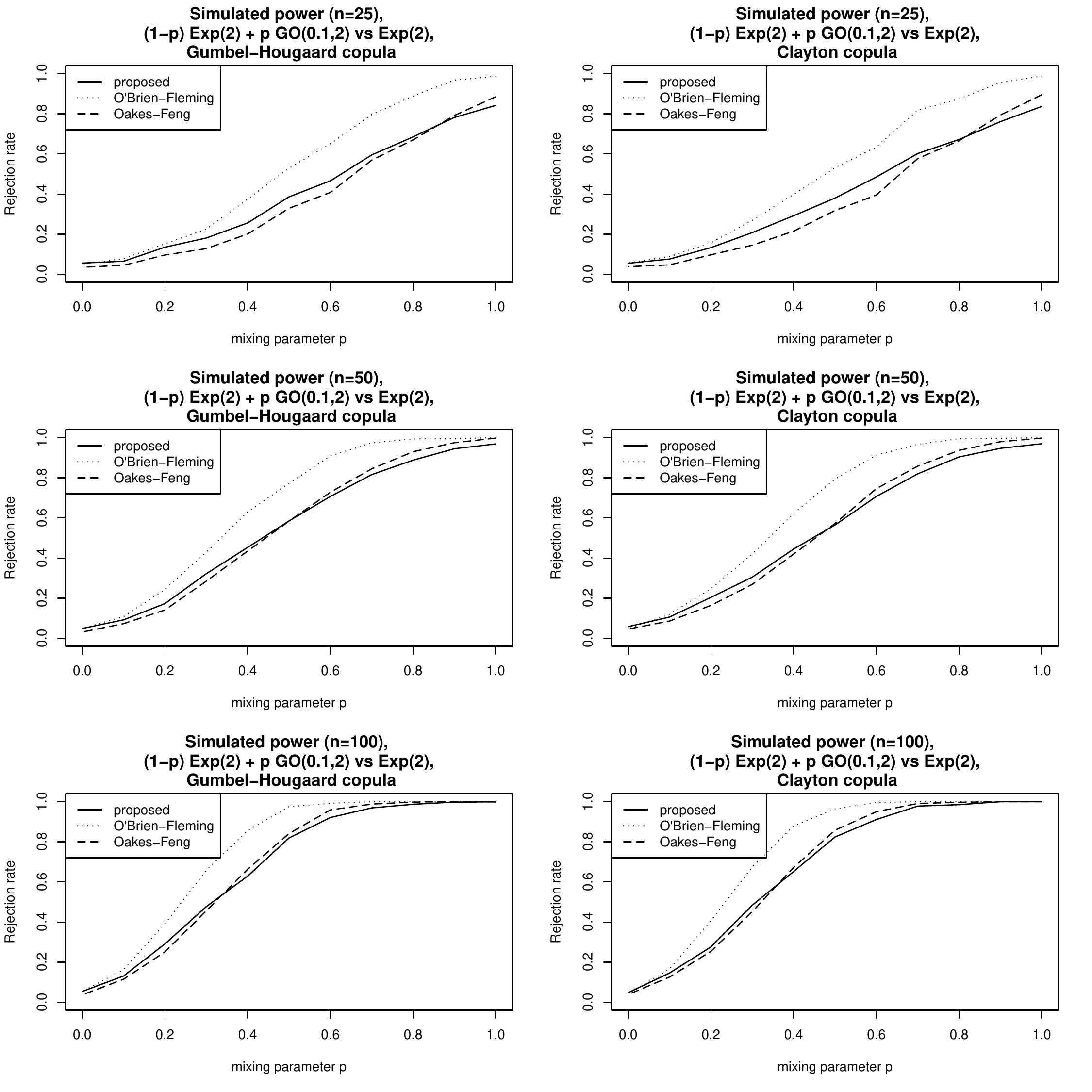}
\caption{Simulated power of three selected right-tailed randomization-based tests with significance level $\alpha=5\%$.}
\label{fig:power_mix2_exp}
\end{figure}

Figures~\ref{fig:power_exp_exp} and~\ref{fig:power_mix2_exp} present the power simulation results of the proposed randomization-based one-sided test in comparison to the paired Prentice-Wilcoxon test and the stratified log-rank test.
Here, the marginal distributions are different competing exponential distributions and exponential-Gompertz mixtures versus an exponential distribution, respectively.
Since also here the overall findings are similar to those given in Figure~\ref{fig:power_mix1_exp} in the main manuscript, no further comments are needed.

\begin{sidewaystable}%[hb]
%\vspace{12cm}
\caption{Simulated sizes of the right-tailed tests (in \%) with nominal significance level $\alpha=1\%$.\\
	Abbreviations: Copula: GH = Gumbel-Hougaard; critical values: asy.\ = asymptotical normal, bs.\ = bootstrap, rand.\ = randomization; tests: lin.\ = linear, tra.\ = $\log-\log$-transformed.}
\centering
\vspace{0.2cm}
%\vspace{16cm}
%\small 
\begin{tabular}{|ccc|cc|cc|cc|cc|cc|cc|cc|cc|cc|}
	\hline
	&&&\multicolumn{6}{c|}{light censoring}&\multicolumn{6}{c|}{medium censoring}&\multicolumn{6}{c|}{strong censoring}\tabularnewline
	&&&\multicolumn{2}{c}{asy.}&\multicolumn{2}{c}{bs.}&\multicolumn{2}{c|}{rand.}&\multicolumn{2}{c}{asy.}&\multicolumn{2}{c}{bs.}&\multicolumn{2}{c|}{rand.}&\multicolumn{2}{c}{asy.}&\multicolumn{2}{c}{bs.}&\multicolumn{2}{c|}{rand.}\tabularnewline
	copula & distribution & $n$ & lin. & tra. &lin. & tra. &lin. & tra. &lin. & tra. &lin. & tra. &lin. & tra. &lin. & tra. &lin. & tra. &lin. & tra.  \tabularnewline
	\hline
	GH&Exm vs.&$ 25$&$2.5$&$0.4$&$0.3$&$2.2$&$1.2$&$1.2$&$3.2$&$0.6$&$1.2$&$2.9$&$1.5$&$1.4$&$3.2$&$0.6$&$1.5$&$2.5$&$1.3$&$1.3$\tabularnewline
	&Exp mix&$ 50$&$1.6$&$0.6$&$0.5$&$1.4$&$1.1$&$1.1$&$2.1$&$0.9$&$0.9$&$1.9$&$1.4$&$1.4$&$2.7$&$1.3$&$1.5$&$2.6$&$1.4$&$1.4$\tabularnewline
	&&$ 75$&$1.3$&$0.6$&$0.5$&$1.0$&$0.9$&$0.9$&$1.9$&$0.7$&$0.8$&$1.6$&$1.3$&$1.3$&$2.1$&$1.3$&$1.3$&$1.8$&$1.5$&$1.5$\tabularnewline
	&&$100$&$1.5$&$0.9$&$1.0$&$1.3$&$1.3$&$1.3$&$1.4$&$0.6$&$0.7$&$1.1$&$0.9$&$0.9$&$2.4$&$1.5$&$1.2$&$1.7$&$1.6$&$1.6$\tabularnewline
	&&$125$&$1.1$&$0.6$&$0.7$&$1.0$&$1.0$&$1.0$&$1.3$&$0.8$&$0.9$&$1.2$&$1.0$&$1.0$&$2.4$&$1.8$&$1.4$&$1.8$&$1.8$&$1.8$\tabularnewline
	&&$150$&$1.4$&$0.9$&$1.1$&$1.2$&$1.3$&$1.3$&$1.2$&$0.9$&$0.9$&$1.0$&$1.1$&$1.1$&$2.1$&$1.5$&$1.4$&$1.7$&$1.3$&$1.2$\tabularnewline\hline
	GH&Gompertz&$ 25$&$2.7$&$0.9$&$0.2$&$0.7$&$1.6$&$1.7$&$4.6$&$1.3$&$0.9$&$1.3$&$2.9$&$3.1$&$6.7$&$2.1$&$2.4$&$1.3$&$5.9$&$5.9$\tabularnewline
	&vs.\ Exp&$ 50$&$1.5$&$0.6$&$0.1$&$0.9$&$1.1$&$1.1$&$2.0$&$1.0$&$0.5$&$1.1$&$1.9$&$1.8$&$3.8$&$2.0$&$0.9$&$2.3$&$5.1$&$5.0$\tabularnewline
	&&$ 75$&$1.3$&$0.6$&$0.5$&$1.0$&$1.0$&$1.0$&$1.4$&$0.7$&$0.4$&$0.9$&$1.4$&$1.4$&$3.1$&$1.8$&$0.8$&$1.7$&$4.7$&$4.6$\tabularnewline
	&&$100$&$1.3$&$0.6$&$0.7$&$1.1$&$1.0$&$1.0$&$1.6$&$0.8$&$0.5$&$0.9$&$1.3$&$1.4$&$1.7$&$1.0$&$0.4$&$0.9$&$3.7$&$3.7$\tabularnewline
	&&$125$&$1.4$&$0.6$&$0.9$&$1.2$&$1.1$&$1.1$&$1.2$&$0.6$&$0.6$&$0.8$&$1.0$&$1.0$&$1.7$&$0.8$&$0.4$&$0.7$&$3.9$&$3.8$\tabularnewline
	&&$150$&$1.4$&$0.9$&$1.1$&$1.4$&$1.4$&$1.4$&$1.2$&$0.6$&$0.7$&$0.9$&$1.0$&$1.0$&$1.5$&$0.8$&$0.5$&$0.7$&$3.0$&$3.0$\tabularnewline\hline
	Clayton&Exp vs.&$ 25$&$2.2$&$0.5$&$0.1$&$2.4$&$1.1$&$1.2$&$2.2$&$0.4$&$0.4$&$2.6$&$1.0$&$0.9$&$2.5$&$0.4$&$0.6$&$2.6$&$1.0$&$1.0$\tabularnewline
	&Exp mix&$ 50$&$1.6$&$0.7$&$0.4$&$1.6$&$1.2$&$1.2$&$1.7$&$0.4$&$0.5$&$1.4$&$0.9$&$0.9$&$1.9$&$0.8$&$0.8$&$1.9$&$1.1$&$1.1$\tabularnewline
	&&$ 75$&$1.1$&$0.5$&$0.5$&$0.9$&$1.0$&$1.0$&$1.4$&$0.7$&$0.6$&$1.3$&$1.1$&$1.1$&$1.5$&$0.8$&$0.9$&$1.3$&$1.2$&$1.2$\tabularnewline
	&&$100$&$1.6$&$0.9$&$1.0$&$1.4$&$1.4$&$1.4$&$1.4$&$0.9$&$0.7$&$1.1$&$1.2$&$1.2$&$1.3$&$0.7$&$0.7$&$1.1$&$1.1$&$1.1$\tabularnewline
	&&$125$&$1.0$&$0.6$&$0.8$&$0.9$&$0.8$&$0.9$&$1.5$&$0.9$&$0.9$&$1.3$&$1.4$&$1.4$&$1.3$&$0.8$&$0.8$&$1.1$&$1.1$&$1.2$\tabularnewline
	&&$150$&$1.0$&$0.6$&$0.7$&$1.0$&$0.9$&$0.9$&$1.2$&$0.7$&$0.8$&$1.0$&$1.0$&$1.0$&$1.2$&$0.8$&$0.8$&$1.0$&$1.1$&$1.1$\tabularnewline\hline
	Clayton&Gompertz&$ 25$&$2.3$&$0.6$&$0.1$&$2.3$&$1.4$&$1.5$&$2.5$&$0.4$&$0.4$&$2.6$&$1.1$&$1.0$&$3.3$&$0.3$&$0.6$&$3.1$&$1.4$&$1.4$\tabularnewline
	&vs.\ Exp&$ 50$&$1.3$&$0.6$&$0.5$&$1.1$&$0.9$&$0.9$&$2.0$&$0.8$&$0.6$&$1.7$&$1.2$&$1.1$&$1.9$&$0.7$&$0.6$&$1.8$&$1.1$&$1.1$\tabularnewline
	&&$ 75$&$1.0$&$0.6$&$0.4$&$0.7$&$0.8$&$0.8$&$1.6$&$0.8$&$0.8$&$1.3$&$1.1$&$1.1$&$1.5$&$0.8$&$0.7$&$1.2$&$1.1$&$1.1$\tabularnewline
	&&$100$&$1.4$&$0.8$&$0.8$&$1.2$&$1.1$&$1.1$&$1.4$&$0.8$&$0.8$&$1.0$&$1.1$&$1.1$&$1.7$&$1.0$&$0.9$&$1.3$&$1.4$&$1.4$\tabularnewline
	&&$125$&$1.0$&$0.7$&$0.7$&$1.0$&$1.0$&$1.0$&$1.4$&$0.8$&$0.8$&$1.1$&$1.2$&$1.2$&$1.7$&$1.1$&$1.0$&$1.3$&$1.5$&$1.4$\tabularnewline
	&&$150$&$1.3$&$0.8$&$0.8$&$1.0$&$1.0$&$1.0$&$1.4$&$0.8$&$0.8$&$1.0$&$1.2$&$1.2$&$1.5$&$1.2$&$1.0$&$1.3$&$1.4$&$1.4$\tabularnewline
	\hline
\end{tabular}
\vspace{0.3cm}
\label{tab:null_1}
\end{sidewaystable}

\begin{sidewaystable}
%\vspace{12cm}
%\hspace{2cm}
\caption{Simulated sizes of the right-tailed tests (in \%) with nominal significance level $\alpha=10\%$. \\
	Abbreviations: Copula: GH = Gumbel-Hougaard; critical values: asy.\ = asymptotical normal, bs.\ = bootstrap, rand.\ = randomization; tests: lin.\ = linear, tra.\ = $\log-\log$-transformed.}
\centering
\vspace{0.2cm}
\begin{tabular}{|ccc|cc|cc|cc|cc|cc|cc|cc|cc|cc|}
	\hline
	&&&\multicolumn{6}{c|}{light censoring}&\multicolumn{6}{c|}{medium censoring}&\multicolumn{6}{c|}{strong censoring}\tabularnewline
	&&&\multicolumn{2}{c}{asy.}&\multicolumn{2}{c}{bs.}&\multicolumn{2}{c|}{rand.}&\multicolumn{2}{c}{asy.}&\multicolumn{2}{c}{bs.}&\multicolumn{2}{c|}{rand.}&\multicolumn{2}{c}{asy.}&\multicolumn{2}{c}{bs.}&\multicolumn{2}{c|}{rand.}\tabularnewline
	copula & distribution & $n$ & lin. & tra. &lin. & tra. &lin. & tra. &lin. & tra. &lin. & tra. &lin. & tra. &lin. & tra. &lin. & tra. &lin. & tra.  \tabularnewline
	\hline
	GH&Exp vs.&$ 25$&$12.3$&$10.1$&$ 8.8$&$10.4$&$10.7$&$10.6$&$11.1$&$ 9.2$&$ 7.8$&$ 9.4$&$10.5$&$10.5$&$ 9.1$&$ 7.9$&$5.4$&$ 6.7$&$12.8$&$12.7$\tabularnewline
	&Exp mix&$ 50$&$10.9$&$ 9.6$&$ 9.4$&$10.0$&$10.4$&$10.4$&$10.4$&$ 9.2$&$ 8.9$&$ 9.6$&$10.3$&$10.2$&$ 7.2$&$ 6.8$&$4.7$&$ 5.2$&$12.4$&$12.4$\tabularnewline
	&&$ 75$&$10.5$&$ 9.4$&$ 9.5$&$10.2$&$10.1$&$10.1$&$11.0$&$ 9.8$&$10.0$&$10.6$&$10.7$&$10.7$&$ 6.8$&$ 6.1$&$4.9$&$ 5.1$&$12.3$&$12.3$\tabularnewline
	&&$100$&$10.7$&$ 9.9$&$10.1$&$10.5$&$10.4$&$10.4$&$10.6$&$ 9.6$&$ 9.7$&$10.1$&$10.1$&$10.1$&$ 7.0$&$ 6.6$&$5.3$&$ 5.5$&$12.3$&$12.3$\tabularnewline
	&&$125$&$ 9.8$&$ 8.9$&$ 9.1$&$ 9.4$&$ 9.6$&$ 9.6$&$10.1$&$ 9.3$&$ 9.6$&$ 9.8$&$ 9.6$&$ 9.6$&$ 6.8$&$ 6.3$&$5.4$&$ 5.6$&$12.8$&$12.8$\tabularnewline
	&&$150$&$10.7$&$ 9.9$&$10.4$&$10.6$&$10.5$&$10.5$&$10.4$&$ 9.6$&$ 9.8$&$10.1$&$ 9.9$&$ 9.9$&$ 6.7$&$ 6.2$&$5.7$&$ 5.7$&$12.7$&$12.7$\tabularnewline\hline
	GH&Gompertz&$ 25$&$13.3$&$11.0$&$ 7.1$&$ 9.4$&$12.0$&$12.0$&$14.2$&$12.0$&$ 6.7$&$ 9.4$&$14.6$&$14.8$&$16.0$&$13.9$&$6.6$&$ 9.3$&$22.9$&$22.7$\tabularnewline
	&vs.\ Exp&$ 50$&$11.2$&$ 9.7$&$ 8.9$&$ 9.8$&$10.5$&$10.5$&$10.8$&$ 9.2$&$ 6.6$&$ 7.4$&$10.9$&$10.9$&$12.2$&$10.9$&$5.4$&$ 7.0$&$19.3$&$19.3$\tabularnewline
	&&$ 75$&$ 9.7$&$ 8.4$&$ 8.5$&$ 8.9$&$ 9.1$&$ 9.2$&$10.2$&$ 8.9$&$ 7.7$&$ 8.6$&$10.1$&$10.1$&$10.5$&$ 9.5$&$5.8$&$ 6.5$&$18.6$&$18.6$\tabularnewline
	&&$100$&$10.5$&$ 9.4$&$ 9.6$&$10.0$&$10.1$&$10.2$&$10.1$&$ 9.0$&$ 8.6$&$ 9.1$&$10.3$&$10.3$&$ 9.0$&$ 8.3$&$5.1$&$ 5.7$&$15.1$&$15.2$\tabularnewline
	&&$125$&$10.8$&$ 9.7$&$10.1$&$10.4$&$10.5$&$10.5$&$10.4$&$ 9.2$&$ 9.5$&$ 9.9$&$10.1$&$10.2$&$ 8.7$&$ 8.1$&$5.5$&$ 5.8$&$15.4$&$15.4$\tabularnewline
	&&$150$&$10.0$&$ 9.4$&$ 9.5$&$ 9.8$&$ 9.8$&$ 9.8$&$ 9.8$&$ 9.0$&$ 9.4$&$ 9.7$&$ 9.7$&$ 9.7$&$ 8.1$&$ 7.3$&$5.3$&$ 5.7$&$14.9$&$14.8$\tabularnewline\hline
	Clayton&Exp vs.&$ 25$&$11.1$&$ 8.9$&$ 7.9$&$ 9.3$&$ 9.8$&$ 9.8$&$11.2$&$ 8.9$&$ 7.1$&$ 8.9$&$ 9.9$&$ 9.9$&$10.9$&$ 8.7$&$7.1$&$ 9.0$&$ 9.9$&$ 9.9$\tabularnewline
	&Exp mix&$ 50$&$ 9.9$&$ 8.6$&$ 8.3$&$ 9.0$&$ 9.6$&$ 9.6$&$10.6$&$ 9.1$&$ 8.6$&$ 9.4$&$10.0$&$10.0$&$ 9.6$&$ 8.6$&$7.0$&$ 7.7$&$ 9.9$&$ 9.9$\tabularnewline
	&&$ 75$&$10.6$&$ 9.3$&$ 9.4$&$ 9.9$&$10.2$&$10.2$&$ 9.2$&$ 8.1$&$ 7.6$&$ 8.2$&$ 9.2$&$ 9.3$&$ 9.7$&$ 8.6$&$7.5$&$ 8.1$&$10.3$&$10.3$\tabularnewline
	&&$100$&$10.8$&$10.0$&$10.1$&$10.5$&$10.7$&$10.7$&$ 9.4$&$ 8.5$&$ 8.6$&$ 8.9$&$ 9.7$&$ 9.7$&$ 9.0$&$ 8.1$&$7.4$&$ 7.8$&$10.1$&$10.1$\tabularnewline
	&&$125$&$ 9.5$&$ 8.8$&$ 9.1$&$ 9.5$&$ 9.6$&$ 9.6$&$10.0$&$ 9.2$&$ 9.0$&$ 9.3$&$10.3$&$10.3$&$ 9.7$&$ 8.9$&$8.2$&$ 8.4$&$11.1$&$11.1$\tabularnewline
	&&$150$&$10.2$&$ 9.4$&$ 9.7$&$ 9.9$&$10.2$&$10.2$&$ 9.5$&$ 8.9$&$ 8.8$&$ 9.0$&$10.3$&$10.3$&$ 9.1$&$ 8.5$&$7.8$&$ 8.1$&$10.6$&$10.6$\tabularnewline\hline
	Clayton&Gompertz&$ 25$&$12.1$&$ 9.6$&$ 7.9$&$ 9.5$&$10.1$&$10.2$&$12.2$&$ 9.7$&$ 7.4$&$ 9.6$&$10.8$&$10.8$&$12.6$&$10.3$&$7.7$&$10.1$&$11.4$&$11.4$\tabularnewline
	&vs.\ Exp&$ 50$&$11.0$&$ 9.5$&$ 8.7$&$ 9.5$&$10.0$&$10.0$&$11.9$&$10.4$&$ 9.3$&$10.2$&$11.0$&$11.0$&$11.3$&$ 9.6$&$8.5$&$ 9.2$&$11.2$&$11.3$\tabularnewline
	&&$ 75$&$10.1$&$ 8.9$&$ 8.7$&$ 9.2$&$ 9.4$&$ 9.4$&$11.2$&$10.1$&$ 9.3$&$ 9.9$&$11.1$&$11.1$&$10.6$&$ 9.5$&$8.4$&$ 9.1$&$11.4$&$11.5$\tabularnewline
	&&$100$&$10.7$&$ 9.7$&$ 9.7$&$10.1$&$10.4$&$10.4$&$10.9$&$ 9.9$&$ 9.3$&$ 9.8$&$10.5$&$10.5$&$10.8$&$ 9.7$&$8.9$&$ 9.3$&$11.3$&$11.3$\tabularnewline
	&&$125$&$ 9.6$&$ 8.7$&$ 8.9$&$ 9.1$&$ 9.4$&$ 9.4$&$ 9.3$&$ 8.6$&$ 8.2$&$ 8.5$&$ 9.4$&$ 9.4$&$10.4$&$ 9.6$&$8.5$&$ 8.9$&$10.8$&$10.8$\tabularnewline
	&&$150$&$10.4$&$ 9.6$&$ 9.5$&$ 9.8$&$10.1$&$10.1$&$11.0$&$10.2$&$ 9.8$&$10.0$&$10.8$&$10.8$&$10.1$&$ 9.6$&$8.7$&$ 9.0$&$11.0$&$11.0$\tabularnewline
	\hline
\end{tabular}
\vspace{0.3cm}
\label{tab:null_10}
\end{sidewaystable}

%\newpage

%\appendixthree
%\section*{Appendix 3}

%{Achtung, der appendix muss vor die references, daher hier umsortiert und auch umstrukturiert (wie im styleguide). Ich würde ggf. überlegen ob die additional simulations in ein supplement kommen? Da muesste man mal schauen wie das sonst so ueblich ist.}
% citations won't show in text... compile offline before handing in!
% unfortunately, the following doesn't work in Overleaf... but it works offline!

\end{document}